\documentclass[journal]{IEEEtran}
\ifCLASSINFOpdf
\else
   \usepackage[dvips]{graphicx}
\fi
\usepackage{url}
\usepackage{graphicx}
\usepackage{amsmath,bm}
\usepackage{amssymb}  
\usepackage{bbm}       
\usepackage{cite}        
 \usepackage{epstopdf} 
  \usepackage{xcolor} 
  \usepackage{subfigure} 
\usepackage{amsthm}    
\usepackage{booktabs} 
\usepackage{colortbl}  
\usepackage{enumitem}  
\usepackage{makecell}  
\usepackage{float}
\usepackage{dsfont}

\usepackage[ruled]{algorithm2e}  
\usepackage{array}
\newcommand{\PreserveBackslash}[1]{\let\temp=\\#1\let\\=\temp}
\newcolumntype{C}[1]{>{\PreserveBackslash\centering}p{#1}}
\newcolumntype{R}[1]{>{\PreserveBackslash\raggedleft}p{#1}}
\newcolumntype{L}[1]{>{\PreserveBackslash\raggedright}p{#1}}
\newcommand{\tabincell}[2]{\begin{tabular}{@{}#1@{}}#2\end{tabular}}

\newfloat{figtab}{htb}{fgtb}
\makeatletter
  \newcommand\figcaption{\def\@captype{figure}\caption}
  \newcommand\tabcaption{\def\@captype{table}\caption}


\usepackage{setspace}

\begin{document}

\title{Deep Multimodal Learning: Merging Sensory Data for Massive MIMO Channel Prediction}

\author{Yuwen Yang,  Feifei Gao,  Chengwen Xing,   Jianping An,   and Ahmed Alkhateeb
\thanks{Y. Yang and F. Gao are with  Institute for Artificial Intelligence Tsinghua University
(THUAI), State Key Lab of Intelligent Technologies and Systems, Beijing National Research Center for Information Science and
Technology (BNRist), Department of Automation, Tsinghua University, Beijing,
100084, P. R. China (email: yyw18@mails.tsinghua.edu.cn, feifeigao@ieee.org).}
\thanks{C. Xing and J. An are with the School of Information and Electronics, Beijing Institute of Technology, Beijing 100081, China
(e-mail: chengwenxing@ieee.org; an@bit.edu.cn).}
\thanks{A. Alkhateeb is with the School of Electrical, Computer and
Energy Engineering at Arizona State University, Tempe, AZ 85287, USA (e-mail: alkhateeb@asu.edu). }
}

\markboth{accepted by IEEE Journal on Selected Areas in Communications}
{Shell \MakeLowercase{\textit{et al.}}: xxxx}
\maketitle

\begin{abstract}
Existing work in  intelligent communications has recently  made preliminary attempts  to
utilize multi-source sensing information (MSI) to improve the system performance. However, the research on MSI aided intelligent communications has not yet explored how to integrate and fuse the multimodal sensory data, which motivates us to develop a systematic framework for wireless communications based on deep multimodal learning (DML).
In this paper, we first present complete descriptions and heuristic understandings on the framework of DML based wireless communications, where core design choices are analyzed in the view of communications. Then, we develop several DML based architectures for channel prediction in massive  multiple-input multiple-output (MIMO) systems that leverage various  modality combinations and fusion levels. The case study of massive MIMO channel prediction offers
an important example that can be followed in developing
other DML based communication technologies. Simulation results demonstrate that the proposed DML framework can effectively exploit the constructive and complementary information of multimodal sensory data  to assist the current wireless communications.
\end{abstract}


\IEEEpeerreviewmaketitle
\addtolength{\topmargin}{+0.1cm} 

\section{Introduction}
To satisfy the demand of explosive wireless applications, e.g., diverse intelligent terminal access, autonomous driving, and Internet of Things, etc, the new generation of wireless communication systems is expected to handle massive data  and meet the requirements of both high-reliability and low-latency. However, existing communication systems, which are basically designed based on conventional communication theories, exhibit several inherent limitations in meeting the aforementioned requirements, such as  relying on accurate theoretical  models, suffering from high complexity algorithms, and being  restricted to block-structure communication protocols, etc \cite{8233654,9113273,8715338}.
Recently, intelligence communication has been recognized as a promising direction in future wireless communications.
 As a major branch of machine learning, deep learning (DL) has been applied in physical layer communications as a  potential  solution to deal with the massive data and the high complexity of wireless communication systems \cite{8663966}.
By merging  DL  into existing communication systems, many remarkable progresses have been made in various applications such as channel estimation
\cite{8922743,8353153,8752012,8672767,8491068,8509622,8979256,9175003,alrabeiah2019deep,8795533,8896030,8815888,8647328},  data detection \cite{8052521,9018199}, channel feedback \cite{8482358,guo2019convolutional},  beamforming 
\cite{alrabeiah2019deep2,weihua20192d,alrabeiah2019viwi}, and hybrid precoding \cite{Li2019}, etc.
In particular, conventional massive multiple-input multiple-output (MIMO)   channel estimation algorithms have hit bottlenecks  due to  the prohibitively high overheads  associated with  massive antennas \cite{8354789,5595728,7801046,8094949}. While DL based massive MIMO channel prediction can significantly improve the prediction accuracy  or reduce the system overheads \cite{8979256,9175003,alrabeiah2019deep,8795533,8896030,8815888,8647328}.

Compared with conventional communications that are based on statistics and information theories,  DL  based communications  benefit from both
the excellent  learning capability of deep neural networks (DNNs) and the impressive computational throughput of parallel processing architectures. Besides, the most advantageous aspect of  DL  is its ability to handle problems with imperfect models or without mathematical models. Interestingly, there exists out-of-band side-information in communication systems that can be utilized to improve the system performance, including sub-6G channels,  user positions, and 3D scene images obtained by cameras, etc. It should be noted that  conventional communication techniques  can hardly take advantage of these out-of-band side-information due to the lack of tractable mathematical models.
In fact, the development of techniques that utilize  out-of-band side-information to improve the system performance has been an emerging research trend in
DL based communications.
For example, \cite{alrabeiah2019deep2} proposed  a sub-6GHz channel information aided network for  mmwave beam and blockage prediction, which could effectively reduce the overheads of both feedback and beam training. In \cite{weihua20192d},
a novel 3D scene based beam selection architecture was  developed  for mmwave communications by using the surrounding 3D scene of the cellular coverage as the network inputs.

Meanwhile, model  aided  DL  has also made much progress  recently.
Instead of purely relying on training data,
 model  aided  DL  benefits from the guidance of model information and therefore can achieve better performance \cite{8715338}. For instance,
 \cite{8509622} proposed an efficient DL  based channel estimator by
 learning the linear model between the least squares (LS) estimation and the linear minimum mean square error (LMMSE) estimation.
In \cite{9018199},  the authors proposed a model-driven DL based MIMO detector by  unfolding an iterative algorithm, which can  significantly outperform the corresponding
traditional iterative detector.

Although a few  DL  based works have  tried to utilize multi-source sensing information (MSI), e.g., out-of-band side-information and model information,  to improve the system performance, none of  them has yet investigated  how to integrate and comprehensively  utilize  MSI in communication systems.
Communication systems naturally work with multimodal data  and this clear advantage should not be squandered.
From the viewpoint  of machine learning, data from various sources are referred as multimodal sensory data, whereas  data from one source are referred as data of a single modality.
Multimodal learning aims to build models that can fully exploit the constructive and complementary information lying in multimodal data, thus gaining performance advantages over methods that only use data of a single modality  \cite{8103116}.
By combining  DL  architectures with multimodal learning methodologies, the concept of deep multimodal learning (DML) has been proposed in \cite{Ngiam11}.
Thanks to the excellent flexibility of  DL
in extracting hierarchical features of  data, DML
offers several advantages over  conventional   multimodal learning, such as learning based feature extraction, implicit dimensionality reduction, and
easily scalability in the modality number, etc \cite{8103116}.
So far, the mainstream applications of DML include human action recognition, audio-visual speaker detection, and autonomous driving, etc \cite{Ngiam11,8103116,7780582}.
For example, \cite{7780582} jointly exploited two modalities, i.e., image and optical flow, for human action recognition, which could obtain higher recognition accuracy than only using image data.

This paper aims to develop a systematic framework on  DML based wireless communications.
By  using DML, the multimodal sensory data available in wireless communication systems can be fully exploited to provide
constructive and complementary information for various tasks. The main contributions of this work can be summarized as following:
\begin{itemize}
  \item We provide complete descriptions and  heuristic analyses on the framework of DML based wireless communications. As opposed to   \cite{8103116} and \cite{Ngiam11} that mainly study DML in computer vision, speech, and natural language processing areas, this is the first work that explores how DML technologies  can be applied to wireless communications to the best of the authors' knowledge.
  \item By investigating various modality combinations and fusion levels, we design several DML based architectures for channel prediction in massive MIMO systems as a case study. The  design process presents  a beneficial guidance on developing DML based communication technologies, while the  proposed architectures can be easily extended to other communication problems like  beam prediction, data detection, and antenna selection.
  \item Simulations based on ray-tracing software have been conducted and  demonstrate that the proposed framework  can effectively exploit the constructive and complementary information of multimodal sensory data in various scenarios.
\end{itemize}

The remainder of this paper is organized as follows.
The framework of DML based wireless communications is given in Section~\ref{secmoti}.
 As a case study,
Section~\ref{secdowm} proposes several DML based architectures for channel prediction in massive MIMO systems.
Numerical results are provided in Section \ref{secsimu}, followed by our  main conclusions  in Section \ref{secconcul}.

\emph{Notation:}
The bold  letters denote vectors or matrices.
 The notation $\textrm{len}(\bm z)$ denotes  the length of the vector $\bm x$.
 The notations $(\cdot)^T$ and  $(\cdot)^H$ respectively denote the transpose and  the conjugate transpose  of a matrix or a vector.
 The notation ${\mathds{C}^{m\times n}}$ represents the $m\times n$ complex vector space.
The notation $\left\| {\bm x} \right\|_2$  denotes the $L_2$  norm of  $\bm x$.
 The notation $ \circ $ represents the composite mapping operation.
 The notations $\Re[\cdot]$ and $\Im[\cdot]$, respectively,  denote  the real and imaginary parts of  matrices, vectors or  scales.
The notations $*$ and $\otimes$  respectively represent the convolution
operation and the matrix element-wise product.
The notation $ E [\cdot]$ represents the expectation with respect to all random variables within the brackets.


\section{DML for Wireless Communications}\label{secmoti}
MSI in communication systems, including out-of-band side-information, model information, and other system information,  is referred as multimodality. Information from one source is  referred as one modality.  

The  framework of DML consists of  three  core parts: selection, extraction  and fusion. The selection process is to select appropriate models as well as  effective modalities for a certain task.
The extraction process is to extract information from involved modalities. The fusion process is to fuse the extracted information in order to obtain a fused representation of the multimodal sensory data.
To embrace the performance advantages offered by DML, there are several design  choices to be considered, including the selections of models, modalities, fusion levels, and fusion strategies, as will be illustrated in the following.

\subsection{Model Selection}\label{discrina}
 DL  models can generally be divided into two categories: discriminative and generative  models. 
Discriminative models aim to learn  the mapping function from the inputs to the outputs, and are typically  used to solve regression and classification tasks. In other words,  given the input $\bm x$ and the label $\bm y$, discriminative  models learn the conditional probability $P(\bm x|\bm y)$ by updating the model parameters. Since the majority of tasks in physical layer communications are to estimate  $\bm y$ based on $\bm x$, such as channel estimation \cite{8672767,8353153,8752012}, data detection \cite{8052521,9018199}, and beamforming
\cite{alrabeiah2019deep2,weihua20192d,alrabeiah2019viwi}, etc, existing DL techniques for physical layer communications mainly  adopt
discriminative models.

Generative models aim to learn the training data distribution that is required to generate new data with similar distributions. More specifically,
generative models learn the joint probability $P(\bm x,\bm y)$ in supervised learning problems, or learn the input data probability $P(\bm x )$ in unsupervised or self-unsupervised learning problems.
For DML problems, generative models
 are useful in the following three aspects: (1) Extracting features from different modalities; (2) Dealing with the situation of   missing modalities during the test stage or lacking labeled data \cite{srivastava2012learning}; (3) Providing  good initialization points for discriminative models, such as  DNNs \cite{6296526}.

\subsection{Modality Selection}
Multimodal sensory data  of communication systems typically have varying \emph{confidence levels}\footnote{The confidence level of one modality refers to the degree of the contribution or the  reliability offered by the modality towards a certain task \cite{Atrey10}.} when accomplishing different tasks. Take the beam selection at the base station (BS) as an example. The
optimal beam vector can be obtained in three ways: (1) Finding the best beam based on known downlink channels \cite{8333702}; (2) Learning from  3D scene images \cite{weihua20192d,alrabeiah2019viwi}; (3) Extrapolating from sub-6G channels \cite{alrabeiah2019deep2}. Among the three modalities,  the downlink channels obviously have higher confidence level than  both the 3D scene images and the sub-6G channels while the  confidence level
of the 3D scene images and the sub-6G channels  depends on specifical scenarios. Nevertheless, even when we have access to the modality with the highest confidence level, there may exist some modalities that could provide complementary information to further improve the performance or robustness of the single-modality based methods \cite{1230212}.  To better understand this, we can refer to the maximum ratio combining (MRC) \cite{tse2005fundamentals}, which is a widely adopted technology to obtain the combining gains of multi-antenna communication systems. MRC could benefit from the antennas with worse channels, which also provides a revelatory understanding about the gains brought by modalities with relatively lower  confidence levels.

Although different modalities could provide complementary information  for a certain multimodal learning task, too much modalities  may lead to information redundancy and an excessively complex fusion process.
Therefore, it is worthwhile to select the optimal modalities
by comprehensively considering  the performance gain and the fusion complexity.
In the area of computer vision and speech,  the modality selection problem is generally  considered as a tradeoff optimization  \cite{Yi3304,Atrey07,lemmela2008selecting}.
For example, \cite{Yi3304} proposed to select the optimal modalities based on the tradeoff between the feature dimensionality and the modality correlations.
However, the authors did not take the confidence levels of modalities into account, which may miss  modalities with high  confidence levels.
In \cite{Atrey07},  the authors utilized a dynamic programming approach to find the optimal subset of modalities based on the three-fold tradeoff between the performance gain, the overall confidence level of the selected subset, and the cost of the selected subset.
In summary, \cite{Yi3304,Atrey07,lemmela2008selecting}
provide heuristic solutions for  the modality selection problem in the context of multimedia data
while there is hardly any literature studying modality selection for communication systems.
A more direct way  adopted in most of existing DML  works  is to manually select modalities by intuition and experiments.


\subsection{Fusion Level Selection}
\begin{figure*}[!t]
\centering
\includegraphics[width=140mm]{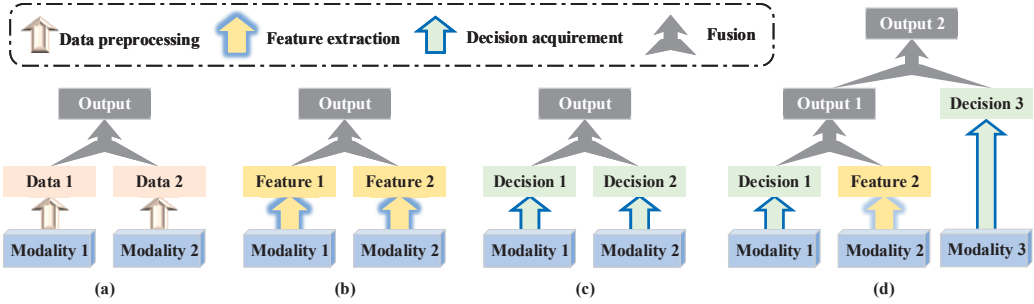}
\caption{Illustrations of various fusion levels for DML. ~(a) data fusion. (b) feature fusion. (c) decision fusion.  (d) hybrid fusion.}
\label{figwhen}
\end{figure*}

\begin{table*}[!t]\footnotesize
  \centering
  \caption{Advantages and disadvantages of various fusion levels}
    \begin{tabular}{|c|l|l|l|l|}
\hline
    \rowcolor[rgb]{ .267,  .447,  .769} \textcolor[rgb]{ 1,  1,  1}{\textbf{Fusion levels}} & \textcolor[rgb]{ 1,  1,  1}{\textbf{Data fusion}} &\textcolor[rgb]{ 1,  1,  1}{\textbf{Feature fusion }} & \textcolor[rgb]{ 1,  1,  1}{\textbf{Decision fusion}} & \textcolor[rgb]{ 1,  1,  1}{\textbf{Hybrid fusion}}  \\
 Disadvantage  &   \makecell[l]{$\cdot$ ignore different modal-\\ \ \ ity structures\\$\cdot $ dimensionality  curse} & \makecell[l]{$\cdot$  additional feature extraction \\\ \ process}   & \makecell[l]{$\cdot$ cannot exploit feature  \\ \ \  level correlations} & $\cdot$   high design
complexity \\
    \hline
    Advantage & $\cdot$ easy to implement & \makecell[l]{$\cdot$ flexible  dimension reduction}      &
    \makecell[l]{$\cdot$ independent decisions\\$\cdot$ easy to fuse} &   \makecell[l]{$\cdot$ independent decisions\\$\cdot$ flexible dimension reduction}  \\
    \hline
    \end{tabular}%
  \label{tabdisadvlevel}%
\end{table*}

 Multimodal sensory data usually have different  dimensionality and structures. For example, in massive MIMO systems, the user position data could be a $3\times 1$ real-valued vector, the received signals could be a much longer complex-valued vector, and the transmitted signals could be a high-dimensional complex-valued  matrix.  To achieve efficient fusion, it is important to select  proper  modality fusion levels.
In general, we can perform the modality fusion in four levels: data fusion, feature fusion, decision  fusion, or hybrid  fusion. 
\subsubsection{Data fusion}
Data fusion is to concatenate the raw or preprocessed data of all the modalities  into a single vector and then to learn a joint multimodal representation based on the concatenated vector during the fusion, as shown in Fig.~\ref{figwhen}~(a).
Data fusion is simple to design and  allows end-to-end training.
An typical example of data fusion can refer to  \cite{8672767}, where the received signals, the transmitted pilots, the pervious channels, and the LS estimates are directly concatenated as the inputs of networks to estimate channels in doubly selective fading scenarios.
However, data fusion ignores the unique structure of  different modalities, which may make it difficult to learn the complementary information among the modalities.
In addition, simple concatenation of the multimodal sensory data  leads to very high dimensional inputs, further resulting in the dimensionality curse problem\footnote{The term ``dimensionality curse'' was first proposed in \cite{bellman1957dynamic}, which refers the phenomenon that when the data dimensionality increases, the dimension of feature space increases so fast that the available data  become sparse and dissimilar in many ways. In this case,
the amount of data required to support the data analysis  often grows exponentially with the dimensionality.}.

\subsubsection{Feature fusion}
Before we introduce the feature fusion, we first explain  how to extract features from the raw  or preprocessed data of one modality.
 The transform from the raw or preprocessed data to features is referred to as ``feature extraction''.
Feature extraction algorithms are either generative or
discriminative, linear or nonlinear, such as principal component analysis, linear discriminative analysis, and Laplacian eigenmaps, etc \cite{belkin2003laplacian}.
In recent few years, DNNs have been recognized as a  popular technique to fuse modalities due to its excellent power and flexibility in extracting hierarchical features  of the data.
Specifically, each  hidden layer of the network indeed represents a  hierarchical features of the inputs.
By changing the number of layers or  choosing proper architecture, DNNs could  extract features at various levels or with various dimensions. For example, \cite{8482358} proposed a deep autoencoder architecture for channel feedback, where the dimension of the learnt compressed vector, i.e., the extracted features that are used to reconstruct the original channel, can be adjusted according to the  given compression ratio.


Now, we discuss the feature fusion. As illustrated in Fig.~\ref{figwhen}~(b), feature fusion is  to fuse
higher-level features   into  a single hidden layer and then to learn a joint multimodal representations for the output.
By utilizing the extracted higher-level features, the model with feature fusion could learn higher-order correlations across modalities.
Moreover, thanks to the  flexibility of feature dimension reduction  offered by DNNs,  the feature fusion strategy may have more advantages than the data fusion strategy in learning multimodal representations \cite{8103116}.

\subsubsection{Decision fusion}
Before we introduce decision fusion, we first explain how to acquire a decision for one modality. The process of obtaining task results based on the modal data is referred to as
 ``decision acquirement''. The decision acquirement can be realized by either  DL  based algorithms or conventional communication algorithms.

As shown in Fig.~\ref{figwhen}~(c), the decisions that are independently acquired by the involved modalities  are fused  to make a final decision, i.e., the output of the model.  The disadvantage of decision fusion  is that it cannot exploit the feature level correlations among modalities.
The decision fusion strategy also has several  advantages over the feature fusion strategy:
\begin{itemize}
  \item When the involved modalities are completely  uncorrelated or have very different dimensionality, it is much simpler and more reasonable to adopt decision fusion.
  \item Decision fusion makes it possible  to adopt the most  suitable  algorithms to make decisions for each modality.  In particular, for the modalities that can use accurate  mathematical models to acquire the decisions, conventional  communication theories based  algorithms would be more suitable than DL based algorithms.
  \item The fusion task would be more easier to implement since the decisions of different modalities usually have  similar data representations.
\end{itemize}
\subsubsection{Hybrid fusion}
To embrace the merits of both the feature  and the
decision fusion strategies, hybrid fusion combines both
feature and decision fusion in a unified framework.
Fig.~\ref{figwhen}~(d) displays an example of hybrid fusion where the decisions and features of three modalities are fused at two different depths of the model.
It should be emphasized  that the decisions or features of multiple modalities can either be fused into a single layer or be fused gradually, i.e., modalities can be fused at different depths of the model. The choice  at what depth to fuse which modalities is based on intuition and experiments.
Take the channel estimation as an example. Given the three modalities, i.e., the pilots, the received signals, and the user position, we usually choose to first fuse the pilots and the  received signals and then fuse the user position because the pilots and the  received signals  are highly correlated and the corresponding fusion should work well based on conventional communication theories.
Besides, the gradual fusion strategy could also avoid overlarge fusion vectors, which
partially solves the problem of dimensionality curse.

Tab.~\ref{tabdisadvlevel} lists the advantages and disadvantages of various fusion levels.
So far, only the data fusion has been utilized in wireless communications, referring to \cite{8672767}.
It should be mentioned that  the fusion level  selection depends on the specifical problem, and therefore the superiority of the fusion level  strategies should be investigated  in a specific problem rather than  in  an absolute sense.

\subsection{Fusion Strategy Selection}
Various methods can be used to fuse different modalities, among which fixed-rule based fusion is the simplest one, including  ``max'', ``min'', ``average'', and ``majority
voting'', etc (see more rules in \cite{667881}).
Besides,  the linear weighted is also a common fusion strategy, where   features or decisions of  different modalities are combined with linear weights.
One successful application of linear weighted is MRC, where the weights can be directly determined by  channels.
However, the linear weighted based modality  fusion is not so simple like MRC. The greatest challenge lies in the determination of the weights for each modality, especially when the data dimension is high.
To solve this problem, DNN based fusion has been proposed and gained growing attentions in these years \cite{Ngiam11,7780582}.  DNN based fusion could learn a nonlinear weighted mapping from the input to the output, and the  weights could be adjusted by  training with pre-acquired datasets instead of manual selection.

\section{Case Study: DML for Massive MIMO Channel Prediction}\label{secdowm}
In this section, we will first  present the available modalities for channel prediction in massive MIMO systems.  We will also give brief descriptions of the involved  network architectures. Then, we will respectively discuss the architecture designs for  the BS and the user, followed
by  detailed  training steps of the proposed networks.

\subsection{Available Modalities for Channel Prediction}
The acquisition of  the channel knowledge plays a critical role in massive MIMO which is a promising technology for future  wireless communication systems  \cite{8354789,7524027}. In this work, we consider a massive MIMO system,
where a BS is equipped with  $M\!\gg\! 1$ antennas in the form of uniform linear array (ULA)\footnote{We adopt the ULA model here for simpler illustration, nevertheless,
the proposed approaches are not restricted  to the specifical array shape, and therefore are applicable for  array with arbitrary geometry.} and serves multiple  single-antenna users. 
Note that the proposed approaches are applicable for uplink/downlink channel prediction in both time division duplex (TDD) and frequency division
duplex (FDD)  systems.
Due to the prohibitively high overheads associated with downlink training and uplink feedback, the downlink channel prediction in FDD massive MIMO system is an  extremely challenging  task \cite{9175003,alrabeiah2019deep,8795533}.
Therefore,  we take  the downlink channel prediction in FDD massive MIMO system as an typical example to illustrate the design and application of DML based channel prediction.

\textbf{System model:} To discuss the available modalities in  FDD massive MIMO systems, we first present the mathematical model for the downlink transmission.
Denote $T_p$ as the pilot length. The received frequency domain signal of the $u$-th user on the $k$-th subcarrier  is
\begin{equation}\label{equreceive}
\bm  r[k]=\bm h_{u}^{T}(f_{\textrm{D},k})\bm s[k] +\bm \varepsilon [k],
\end{equation}
where $\bm  r[k]\in \mathds{C}^{1\times T_p}$ is the received signal, $\bm s[k]\in \mathds{C}^{M\times T_p}$ is the downlink pilot signal,
$\bm \varepsilon[k]$  is the additive white Gaussian noise.
Moreover,  $\bm h_{u}(f_{\textrm{D},k})\in \mathds{C}^{M\times 1}$ is the downlink channel that can be written as \cite{alkhateeb2019deepmimo} 
 \begin{equation}\label{equchannel}
\bm h_{u}(f_{\textrm{D},k})  = \sum\limits_{p = 1}^P {{\alpha _{u,p}}{e^{ - j2\pi f_{\textrm{D},k}\tau_{u,p} + j{\phi _{u,p}}}} \bm a\left( {\theta_{\mathrm{az}}^{u,p}, \theta_{\mathrm{el}}^{u,p}} \right)},
\end{equation}
where $P$ is the path number, $f_{\textrm{D},k}$ is the  frequency of the $k$-th downlink subcarrier, while  $\alpha _{u,p}$, $\phi _{u,p}$, and $\tau _{u,p}$ are the  attenuation, phase shift, and  delay
 of the $p$-th path, respectively. In addition, $\bm a\left( {\theta_{\mathrm{az}}^{u,p}, \theta_{\mathrm{el}}^{u,p}} \right)$ is the array manifold vector defined as
\begin{eqnarray}\label{equavec}
\bm a\left( {\theta_{\mathrm{az}}^{u,p}, \theta_{\mathrm{el}}^{u,p}} \right){\kern -8pt}&={\kern -8pt}&\left[1, e^{j\varpi \sin \left(\theta_{\mathrm{el}}^{u,p}\right) \cos \left(\theta_{\mathrm{az}}^{u,p}\right)}, \ldots\right.  \nonumber \\
& &{\kern 5pt}\left.\ldots, e^{j \varpi\left(M-1\right) \sin \left(\theta_{\mathrm{el}}^{u,p}\right) \cos \left(\theta_{\mathrm{az}}^{u,p}\right)}\right]^{T},
\end{eqnarray}
where  $\varpi ={{2\pi d f_{\textrm{D},k}}}/{c }$, $d$  is  the antenna spacing,  $c$ is the speed of light, and $\{\theta_{\mathrm{az}}^{u,p}, \theta_{\mathrm{el}}^{u,p}\}$  is the \{azimuth, elevation\} angle  of  arrival.
We employ the accurate 3D ray-tracing simulator Wireless  InSite \cite{alkhateeb2019deepmimo} to obtain the channel parameters in Eq.~\eqref{equchannel}, i.e., $\{\alpha _{u,p},\phi _{u,p},\tau _{u,p},\theta_{\mathrm{az}}^{u,p}, \theta_{\mathrm{el}}^{u,p}\}$.
To simplify the notation, we drop the sub-carrier index $k$ and the user index $u$
in the rest of the paper, e.g., replacing  $\bm  r[k], \bm h_{u}(f_{\textrm{D},k})$, and $\bm s[k]$
with $\bm  r, \bm h(f_{\textrm{D}})$, and $\bm s$, respectively.

\textbf{Available Modality:}
 Available  modalities in FDD massive MIMO system could be the received signals, the pilots, the LS estimate, the downlink channels of previous coherent time periods, the uplink channel,  the user location, and the partial downlink channel, as described in the following.
\subsubsection{Received signals and pilots}
Eq.~\eqref{equreceive} indeed reveals that there exists a  mapping function from $\{\bm r,\bm s\}$ to $\bm h(f_{\textrm{D}})$, which  indicates that the received signals $\bm r$ and the pilots $\bm s$ are two modalities that could be jointly utilized to predict the downlink channel  $\bm h(f_{\textrm{D}})$.
\subsubsection{LS estimate}
When the number of pilots are sufficient (i.e., $T_p\!\geq\!M$),   $\bm h(f_{\textrm{D}})$ can be estimated by  LS  \cite{1597555}, i.e.,
\begin{equation}\label{equls}
 \hat{\bm h}_{\mathrm{LS}}(f_{\textrm{D}})=\bm r\bm s^{H} (\bm s \bm s^{H})^{-1}.
\end{equation}
In fact, the LS estimate $ \hat{\bm h}_{\mathrm{LS}}(f_{\textrm{D}})$ can be regarded as one modality from model information.

\subsubsection{Previous downlink channels}
Denote the superscript $(\tau)$  as the index of  coherent time periods.
The downlink channels of previous coherent time periods, i.e., $\bm h^{(\tau-1)}(f_{\textrm{D}}),\bm h^{(\tau-2)}(f_{\textrm{D}}),\cdots$, are referred as previous downlink channels, $\mathord{\buildrel{\lower3pt\hbox{$\scriptscriptstyle\leftarrow$}}
\over {\bm h}} (f_{\textrm{D}})$, for ease of exposition\footnote{Since the downlink channel to be predicted and other modalities involved are all the  data in the $\tau$-th  coherent time period, we have omitted the superscript $(\tau)$ of these realtime data for simplicity.}.
In practical systems, there exist unknown time correlations among channels that cannot be exploited by conventional channel estimation algorithms. Whereas such time correlations could be implicitly learned by DNNs and then be used to  improve the prediction accuracy \cite{8491068,8672767,8979256}.
\subsubsection{User location}
The user location can be obtained by various techniques, such as the ultra-wideband, the global positioning system, and the wireless fidelity, etc. Many positioning works have revealed  that there is a distinct  link between the user's position and channels  \cite{8292280}.
Define the location-to-channel mapping  as
${\bm \Phi _{f}}:\left\{ \bm D_{{\left(x,y,z\right)} }\right\} \to \left\{ \bm h(f )\right\}$,
where ${\bm D_{{\left(x,y,z\right)} }}$ is the 3D coordinate of the user, and $f$ is the carrier frequency.
Based on the universal approximation theorem \cite{hornikmultilayer} and the widely adopted assumption that $\bm \Phi _{f}$ is a bijective deterministic mapping in massive MIMO systems \cite{8292280}, we know that  the mapping function ${\bm \Phi _{f}}$ could be approximated arbitrarily well by a DNN under ideal conditions.
Therefore, the modality of user location could be adopted to predict the downlink channel by using DNNs  to learn the mapping ${\bm \Phi _{f_{\mathrm{D}}}}$.
\subsubsection{Uplink channel}
Since uplink channels are easier to obtain than downlink channels
in massive MIMO systems, many studies  utilize  uplink channels to aid the downlink channel prediction \cite{alrabeiah2019deep,8795533,8334183}.
With the assumption that $\bm \Phi _{f}$ is a bijective deterministic mapping, the  channel-to-location mapping $\bm \Phi _{f}^{-1}$ exists and can be written as
${\bm \Phi _{f}^{-1}}:  \left\{ \bm h(f )\right\}\to \left\{ {{\bm D_{{\left(x,y,z\right)} }}} \right\}$.
Hence, the uplink-to-downlink mapping
$\bm \Psi _{{\textrm{U}} \to {\textrm{D}}}$ exists,
 and can be written as follows \cite{alrabeiah2019deep}:
\begin{eqnarray}
\bm \Psi _{{\textrm{U}} \to {\textrm{D}}} \!\!\!\! &=& \!\!\!\! {\bm \Phi _{f_{\textrm{D}}}}\circ{\bm \Phi _{f_{\textrm{U}}}^{-1}} : \left\{ \bm h(f_{\textrm{U}} )\right\} \to \left\{ \bm h(f_{\textrm{D}} )\right\},
\end{eqnarray}
where $f_{\textrm{U}}$ is the uplink frequency, and ${\bm \Phi _{f_{\textrm{D}}}}\circ{\bm \Phi _{f_{\textrm{U}}}^{-1}} $ represents the
composite mapping related to ${\bm \Phi _{f_{\textrm{D}}}}$ and ${\bm \Phi _{f_{\textrm{U}}}^{-1}} $.
Therefore, the modality of uplink channel could also be adopted to predict the downlink channel by using DNNs  to learn the mapping $\bm \Psi _{{\textrm{U}} \to {\textrm{D}}}$.

\begin{table*}[!t]\footnotesize
  \centering
  \caption{Modalities involved in downlink channel prediction}
    \begin{tabular}{|c|c|c|c|c|c|c|}
\hline
    \rowcolor[rgb]{ .267,  .447,  .769} \textcolor[rgb]{ 1,  1,  1}{\textbf{Modality}} & \textcolor[rgb]{ 1,  1,  1}{\textbf{  $\{\bm r,\bm s\}$}} &\textcolor[rgb]{ 1,  1,  1}{\textbf{  $ \hat{\bm h}_{\mathrm{LS}}(f_{\textrm{D}})$ \cite{8672767,8509622}}} & \textcolor[rgb]{ 1,  1,  1}{\textbf{ $\mathord{\buildrel{\lower3pt\hbox{$\scriptscriptstyle\leftarrow$}}
\over {\bm h}} (f_{\textrm{D}})$\cite{8672767,8491068}}} & \textcolor[rgb]{ 1,  1,  1}{\textbf{${\bm D_{{\left(x,y,z\right)} }}$}} & \textcolor[rgb]{ 1,  1,  1}{\textbf{$\bm h(f_{\textrm{U}} )$\cite{9175003,alrabeiah2019deep,8795533}}} & \textcolor[rgb]{ 1,  1,  1}{\textbf{$\mathord{\buildrel{\lower3pt\hbox{$\scriptscriptstyle\smile$}}
\over  {\bm h}}(f_{\textrm{D}}) $\cite{alrabeiah2019deep,8647328}}} \\
    \rowcolor[rgb]{ .851,  .882,  .949} BS side  &  $ \surd$ & $\times$   & $\surd$    & $\surd$   & $\surd$     & $\surd$ \\
    User side & $\surd$  &$\surd$   & $\surd$   &$\surd$     &  $ \times$     & $\surd$ \\
    \hline
    \end{tabular}%
  \label{tabmodal}%
\end{table*}

\subsubsection{Partial downlink channel}
Due to the high cost and power consumption of the radio-frequency chains, massive MIMO
systems usually adopt  hybrid analog and digital transceivers that are operated with switchers  \cite{Li2019}.
Therefore, given the limited transmission period and pilot length, only
partial downlink channel can be obtained by the user and then be fed back to BS.
Denote  the known partial downlink channel as $\mathord{\buildrel{\lower3pt\hbox{$\scriptscriptstyle\smile$}}
\over  {\bm h}}(f_{\textrm{D}})$  with $\mathrm{len}(\mathord{\buildrel{\lower3pt\hbox{$\scriptscriptstyle\smile$}}
\over  {\bm h}}(f_{\textrm{D}}))<M$.  Denote the vector consisting of unknown elements in $\bm h(f_{\textrm{D}} )$ as $\mathord{\buildrel{\lower3pt\hbox{$\scriptscriptstyle\frown$}}
\over {\bm h}} (f_{\textrm{D}})$.  Recalling Eq.~\eqref{equavec} and  Eq.~\eqref{equchannel}, it is obvious that there exists a  deterministic mapping from $\mathord{\buildrel{\lower3pt\hbox{$\scriptscriptstyle\smile$}}
\over  {\bm h}}(f_{\textrm{D}})$ to $\mathord{\buildrel{\lower3pt\hbox{$\scriptscriptstyle\frown$}}
\over {\bm h}} (f_{\textrm{D}})$, which can be written as
$\bm \Upsilon : \{\mathord{\buildrel{\lower3pt\hbox{$\scriptscriptstyle\smile$}}
\over  {\bm h}}(f_{\textrm{D}})\} \to \{\mathord{\buildrel{\lower3pt\hbox{$\scriptscriptstyle\frown$}}
\over {\bm h}} (f_{\textrm{D}})\}$.
Therefore, we can predict the downlink channel by learning the mapping $\bm \Upsilon$.




In order to facilitate the analysis, we list the  modalities for downlink channel prediction  in  Tab.~\ref{tabmodal},
where ``$\surd$'' and ``$\times$''  respectively represent the available and unavailable modalities for BS or the user.
In particular, the  modalities $\{\bm r,\bm s\}$ and $\mathord{\buildrel{\lower3pt\hbox{$\scriptscriptstyle\smile$}}
\over  {\bm h}}(f_{\textrm{D}}) $ are available for BS because $\bm r$ and $\mathord{\buildrel{\lower3pt\hbox{$\scriptscriptstyle\smile$}}
\over  {\bm h}}(f_{\textrm{D}}) $ could  be fed back to the BS by the user. The modality $\hat{\bm h}_{\mathrm{LS}}(f_{\textrm{D}})$ is obtained based on $\{\bm r,\bm s\}$. When the length of $\bm r$ is sufficiently long for the LS estimator, i.e.,  $T_p\!\geq\!M$, it would be more efficient to directly feed back the downlink channel rather than  $\bm r$   to BS. Therefore, we set the modality $\hat{\bm h}_{\mathrm{LS}}(f_{\textrm{D}})$ to be unavailable at BS.
Tab.~\ref{tabmodal} also displays the existing works that utilize aforementioned modalities to predict channels.
By trying and testing possible modality combinations and  feature level strategies, we can find the modalities with higher confidence levels and the modality combinations with better performance.

\subsection{DNN Architectures}

Based on the definition in Section~\ref{discrina}, the downlink channel prediction is a typical discriminative regression task.
 Since discriminative models are naturally suitable for feature extraction and decision acquirement in discriminative tasks, we choose discriminative models for downlink CSI prediction.
The selections of both modalities and  feature level strategies depends on  specifical scenarios.
Besides, due to the excellent learning capability of DNNs, we adopt DNN based fusion for channel prediction rather than fixed-rule based fusion.

\textbf{Activation function:} The activation functions, including leaky rectified linear units (LeakyReLU)\footnote{We adopt LeakyReLU instead of normal  rectified linear units (ReLU) to avoid the ``dead Relu'' phenomenon \cite{maas2013rectifier}.}, Sigmoid, and Tanh,  apply element-wise nonlinear transformations to the outputs of the network layers. The functions LeakyReLU, Sigmoid, and Tanh can  be respectively written as
  ${\mathcal{ F}}_{\mathrm{LR}}(\bm x) = \max\{\bm x, 0.2  \bm x\}$,
  $\mathcal{F}_{\mathrm{Sig}}(\bm x)= 1/ (1+e^{-\bm x})$, and
  $\mathcal{F}_{\mathrm{Tan}}(\bm x) =(e^{\bm x}-e^{-\bm x})/ (e^{\bm x}+e^{-\bm x})$.

\textbf{Loss function:} A DNN architecture consists of the input $\bm x$, the label $\bm y$, the output $\hat{\bm y}$, the network parameter $\bm \Omega$,  activation functions, the loss function $\textrm{Loss}\left(\bm \Omega\right)$, a back-propagation learning algorithm, and  network layers.  Specifically,  the network parameter $\bm \Omega$ includes the weights and the biases of the network layers. The loss  function adopted in this work is
\begin{equation}\label{equloss}
\textrm{Loss}\left(\bm \Omega\right) =\frac{1}{{V}}\sum\limits_{v= 0}^{ V-1}\left\|\hat{\bm{y}}_{v}-{\bm{y}}_{v} \right\|_{2}^{2},
\end{equation} 
where   $V$ is the batch size, and the subscript $v$ denotes the index of the $v$-th  training sample. The back-propagation learning algorithm adopted in this work is the adaptive moment estimation (ADAM) algorithm \cite{kingmaadam}.
In the off-line training stage, the network parameter $\bm \Omega$ is updated by the ADAM algorithm to  minimize
the loss function $\textrm{Loss}\left(\bm \Omega\right)$ on the training dataset.
 While in the on-line  testing stage,  $\bm \Omega$  is fixed and the network could directly output the  estimates of the labels in the testing dataset with a rather small error.

\begin{figure}[!t]
\centering
\includegraphics[width=80mm]{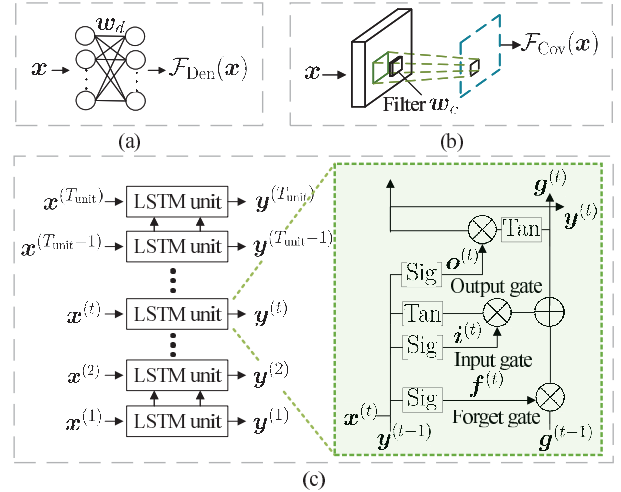}
\caption{Illustrations of the dense layer (a), the convolution layer (b), and the LSTM layer (c).}
\label{figlstm}
\end{figure}

\begin{figure*}[!t]
\centering
\includegraphics[width=160mm]{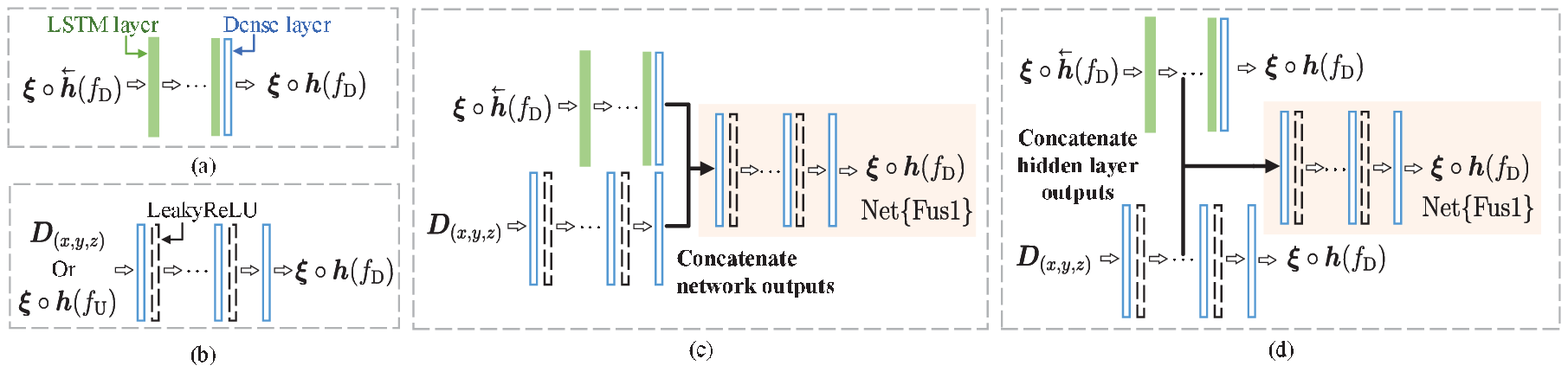}
\caption{The network structures of $\mathrm{Net}{\{\mathtt{H}\}}$ (a),  $\mathrm{Net}{\{\mathtt{L}\}}$ (b), $\mathrm{Net}{\{\mathtt{U}\}}$ (b),
$\mathrm{Net}{\{\mathtt{H},\mathtt{L}\}_{\mathrm{d}}}$ (c), and
$\mathrm{Net}{\{\mathtt{H},\mathtt{L}\}_{\mathrm{f}}}$ (d). }
\label{figbsnoa}
\end{figure*}

\textbf{Network layer:}  Fig.~\ref{figlstm} depicts the structure of the  network layers,  including the dense, the convolution and the LSTM layers. As shown in Fig.~\ref{figlstm}~(a),
the dense layer can be mathematically expressed as
$\mathcal{F}_{\mathrm{Den}} (\bm x)= \bm w_d\bm x+\bm b_d$,
where $\bm w_d$ and $\bm b_d$ are the  weight and the bias of the dense layer,  respectively.
Compared with the  dense layer,
the convolution layer is more powerful in learning the spatial features of the inputs.
As illustrated in Fig.~\ref{figlstm}~(b), the convolution layer can be mathematically expressed as $\mathcal{F}_{\mathrm{Cov}} (\bm x)= \bm w_c*\bm x+\bm b_c$,
where $\bm w_c$ and $\bm b_c$ is the  weight and the bias of the filter, respectively.
Fig.~\ref{figlstm}~(c) depicts the structure of the LSTM layer, where each LSTM layer contains $T_{\mathrm{unit}}$ LSTM units. The output of the LSTM  layer can be written as \[\mathcal{F}_{\mathrm{LSTM}} (\bm x^{(1)},\cdots,\bm x^{(T_{\mathrm{unit}})})=(\bm y^{(1)},\cdots,\bm y^{(T_{\mathrm{unit}})}).\]
 In the $t$-th
($1\le t\le T_{\mathrm{unit}}$) LSTM units,
the relationships between the input $\bm x^{(t)}$ and output $\bm y^{(t)}$  can be expressed with the following equations:
\begin{subequations}
\begin{eqnarray}\label{equlstm}
\bm{i}^{(t)} {\kern -5pt}&=&{\kern -5pt} \mathcal{F}_{\mathrm{Sig}}\left(\bm w_{i} \left[\bm y^{(t-1)}, \bm x^{(t)}\right]+\bm b_{i}\right); \\
\bm{f}^{(t)} {\kern -5pt}&=&{\kern -5pt}\mathcal{F}_{\mathrm{Sig}}\left(\bm w_{f} \left[\bm y^{(t-1)}, \bm x^{(t)}\right]+\bm b_{f}\right); \\
\bm{o}^{(t)} {\kern -5pt}&=&{\kern -5pt}\mathcal{F}_{\mathrm{Sig}}\left(\bm w_{o} \left[\bm y^{(t-1)}, \bm x^{(t)}\right]+\bm b_{o}\right); \\
\bm{g}^{(t)} {\kern -5pt}&=&{\kern -5pt}\bm f^{(t)} \otimes \bm g^{(t-1)}+\nonumber \\ & &{\kern -5pt}\bm i^{(t)} \otimes \mathcal{F}_{\mathrm{Tan}} \left(\bm w_{g }\left[\bm y^{(t-1)}, \bm x^{(t)}\right]+\bm b_{g}\right); \\
\bm{y}^{(t)} {\kern -5pt}&=&{\kern -5pt}\bm o^{(t)}\otimes \mathcal{F}_{\mathrm{Tan}}  \left(\bm g^{(t)}\right), \label{equlstmlast}
\end{eqnarray}
\end{subequations}
where $\{\bm w_{i},\bm w_{f},\bm w_{o},\bm w_{g}\} $ and
$\{\bm b_{i},\bm b_{f},\bm b_{o},\bm b_{g}\}$ are respectively the weights and the biases of the LSTM units, while $\bm{i}^{(t)} $, $\bm{f}^{(t)} $ and $\bm o^{(t)}$ are respectively the input gate, the forget gate and output gate. Moreover,  $\bm g^{(t)}$ is the cell state of the $t$-th LSTM unit.
Since  the LSTM layer  can effectively learn both the short-term and  the long-term features through the
memory cell and the gates, it has been recognized as
a useful tool for time series related tasks.

\subsection{Architecture Designs at the BS Side}\label{secbs}
Accurate downlink channels are crucial for BS to obtain high beamforming gains. Here we consider the
downlink channel prediction problem under two different scenarios, i.e., feedback link being unavailable or available.
Before coming to specifical architectures, we first present our main idea to design fusion architectures as follow:
\begin{enumerate}[label=(\roman*)]
\item  Design and train elementary networks, i.e., the networks that  adopt as few as possible modalities
 to independently predict downlink channels. In fact, all the modalities listed in Tab.~\ref{tabmodal} can independently predict  downlink channels except the two modalities $\{\bm r,\bm s\}$ that should be
 jointly utilized to obtain downlink channels. Note that the performance of the elementary networks can be used to measure the confidence levels of the corresponding modalities.
\item Design and train two-element based networks, i.e., the networks that fuse two elementary networks. The performance of the two-element based networks can be used to measure the complementarity of the corresponding modality combinations. When we design fusion architectures with  multiple modalities, we will   preferentially fuse the modality combinations with better performance and then fuse the modalities with higher confidence levels based on experiments and intuition \cite{8103116}.
\end{enumerate}
The idea is also applicable to the architecture designs at the user side as will be shown
in the later section.

\subsubsection{Feedback link is unavailable}\label{secbsno}
In this scenario, available modalities are the  previous downlink channels $\mathord{\buildrel{\lower3pt\hbox{$\scriptscriptstyle\leftarrow$}}
\over {\bm h}} (f_{\textrm{D}})$, the user location
${\bm D_{{\left(x,y,z\right)} }}$, and the uplink channel $\bm h(f_{\textrm{U}} )$.
To investigate the  confidence levels of the three modalities, we propose three networks, i.e.,   $\mathrm{Net}{\{\mathtt{H}\}}$,
$\mathrm{Net}{\{\mathtt{L}\}}$, and $\mathrm{Net}{\{\mathtt{U}\}}$ to respectively predict the downlink channel based on the  previous downlink channels, the user location, and the uplink channel.
Fig.~\ref{figbsnoa}~(a) illustrates the network structure of $\mathrm{Net}{\{\mathtt{H}\}}$.
The input of $\mathrm{Net}{\{\mathtt{H}\}}$ is $\bm \xi\circ \mathord{\buildrel{\lower3pt\hbox{$\scriptscriptstyle\leftarrow$}}
\over {\bm h}} (f_{\textrm{D}})$, where
\[\mathord{\buildrel{\lower3pt\hbox{$\scriptscriptstyle\leftarrow$}}
\over {\bm h}} (f_{\textrm{D}})=[ \bm h^{(\tau-1)}(f_{\textrm{D}}),\cdots, \bm h^{(\tau-T_{\mathrm{unit}})}(f_{\textrm{D}})],\]
  while $\bm \xi$ is the  mapping between the complex and the real domains, i.e.,
 $\bm \xi: \bm z \to \left({\Re {\left(\bm z^T\right)},\Im {\left(\bm z^T\right)}}\right)^T$.
The label of $\mathrm{Net}{\{\mathtt{H}\}}$ is $\bm \xi\circ  {\bm h}(f_{\textrm{D}})$.
 The network $\mathrm{Net}{\{\mathtt{H}\}}$ is composed of several  LSTM layers and one dense layer.
Here we adopt the LSTM layer to predict the downlink channels for its  superiority in  time
series data analyses.
Besides, we add the dense layer after the last LSTM layer is to release the output of $\mathrm{Net}{\{\mathtt{H}\}}$ from the limited value range of the activation functions  $\mathcal{F}_{\mathrm{Tan}}$ and $\mathcal{F}_{\mathrm{Sig}}$, as indicated in Eq.~\eqref{equlstmlast}.
Fig.~\ref{figbsnoa}~(b) shows the  network structure of both $\mathrm{Net}{\{\mathtt{L}\}}$ and $\mathrm{Net}{\{\mathtt{U}\}}$, where the network is composed of several  dense layers,  and each  dense layer except for the output layer  is followed by the LeakyReLU function.
 Note that $\mathrm{Net}{\{\mathtt{L}\}}$ and $\mathrm{Net}{\{\mathtt{U}\}}$ have the common network structure and the same label $\bm \xi\circ  {\bm h}(f_{\textrm{D}})$, but they have different inputs 
 and
different hype-parameters, including the number of layers, the number of neurons in each layers, and the learning rates, etc.

To investigate the  complementarities of the three modalities i.e., $\mathord{\buildrel{\lower3pt\hbox{$\scriptscriptstyle\leftarrow$}}
\over {\bm h}} (f_{\textrm{D}})$,
${\bm D_{{\left(x,y,z\right)} }}$, and  $\bm h(f_{\textrm{U}} )$, we first propose $\mathrm{Net}{\{\mathtt{L},\mathtt{U}\}_{\mathrm{d}}}$, $\mathrm{Net}{\{\mathtt{H},\mathtt{U}\}_{\mathrm{d}}}$, and
$\mathrm{Net}{\{\mathtt{H},\mathtt{L}\}_{\mathrm{d}}}$
 to respectively fuse two of the three modalities at the decision levels.
As shown in Fig.~\ref{figbsnoa}~(c), $\mathrm{Net}{\{\mathtt{H},\mathtt{L}\}_{\mathrm{d}}}$ consists of $\mathrm{Net}{\{\mathtt{H}\}}$,  $\mathrm{Net}{\{\mathtt{L}\}}$, and $\mathrm{Net}{\{\mathrm{Fus}1\}}$. The network $\mathrm{Net}{\{\mathrm{Fus}1\}}$, composed of several  dense layers  and  LeakyReLU functions, concatenates the network outputs of $\mathrm{Net}{\{\mathtt{H}\}}$ and $\mathrm{Net}{\{\mathtt{L}\}}$ as its input vector.
Note that the structures of $\mathrm{Net}{\{\mathtt{L},\mathtt{U}\}_{\mathrm{d}}}$ and $\mathrm{Net}{\{\mathtt{H},\mathtt{U}\}_{\mathrm{d}}}$ can be similarly  obtained following the design of $\mathrm{Net}{\{\mathtt{H},\mathtt{L}\}_{\mathrm{d}}}$.
 Therefore, we omit the descriptions of these networks for simplicity.
Then, we propose $\mathrm{Net}{\{\mathtt{L},\mathtt{U}\}_{\mathrm{f}}}$, $\mathrm{Net}{\{\mathtt{H},\mathtt{U}\}_{\mathrm{f}}}$, and
$\mathrm{Net}{\{\mathtt{H},\mathtt{L}\}_{\mathrm{f}}}$
 to respectively fuse two of the three modalities at the feature levels.
As shown in Fig.~\ref{figbsnoa}~(d), the main difference between  $\mathrm{Net}{\{\mathtt{H},\mathtt{L}\}_{\mathrm{f}}}$ and
$\mathrm{Net}{\{\mathtt{H},\mathtt{L}\}_{\mathrm{d}}}$
is that $\mathrm{Net}{\{\mathtt{H},\mathtt{L}\}_{\mathrm{f}}}$ concatenates the hidden layer outputs rather than the network outputs of $\mathrm{Net}{\{\mathtt{H}\}}$ and  $\mathrm{Net}{\{\mathtt{L}\}}$.
Similarly, we omit the descriptions of $\mathrm{Net}{\{\mathtt{L},\mathtt{U}\}_{\mathrm{f}}}$ and $\mathrm{Net}{\{\mathtt{H},\mathtt{U}\}{\mathrm{f}}}$
  for simplicity.
 It should be explained that we do not consider  data level fusion for $\mathord{\buildrel{\lower3pt\hbox{$\scriptscriptstyle\leftarrow$}}
\over {\bm h}} (f_{\textrm{D}})$,
${\bm D_{{\left(x,y,z\right)} }}$, and  $\bm h(f_{\textrm{U}} )$ since the three modalities have  very different dimensions and data structures, which would result in inefficient data fusion.

\begin{figure}[!t]
\centering
\includegraphics[width=70 mm]{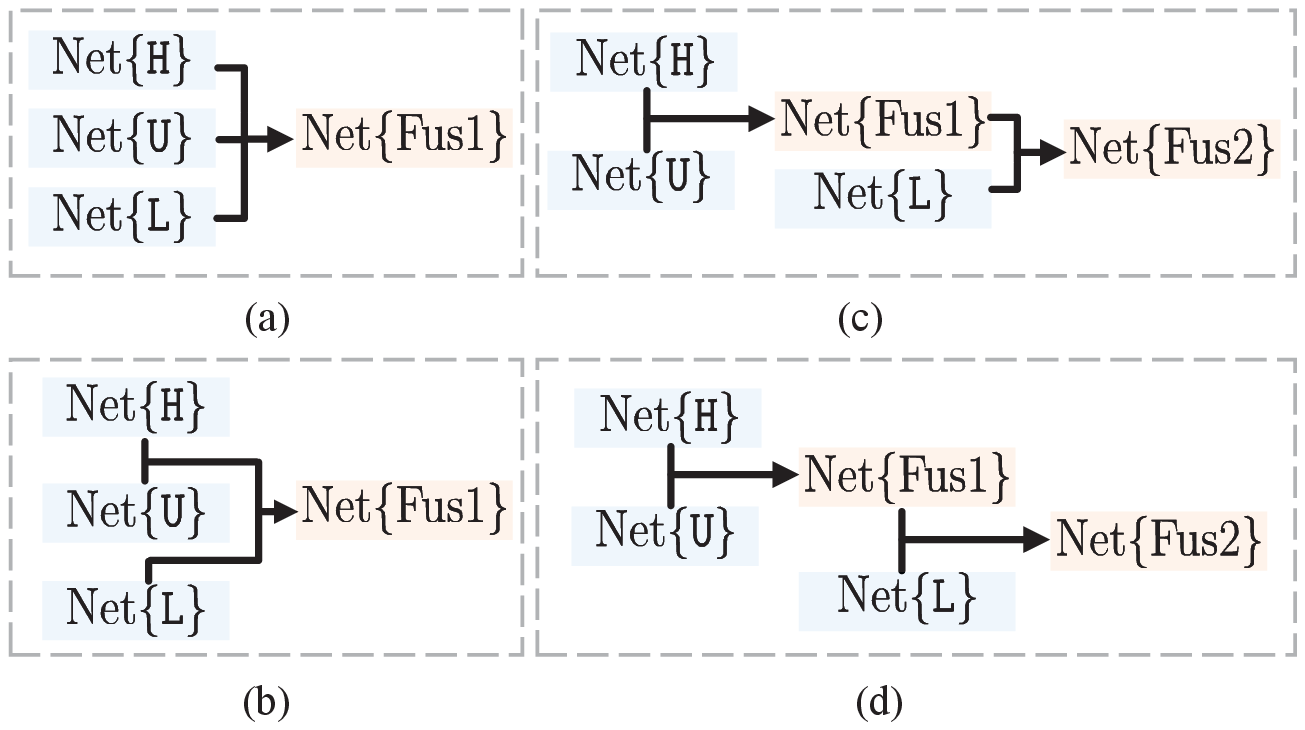}
\caption{The network structures of $\mathrm{Net}{\{\mathtt{H},\mathtt{L}, \mathtt{U}\}_{\mathrm{d}}}$ (a),
$\mathrm{Net}{\{\mathtt{H},\mathtt{L}, \mathtt{U}\}_{\mathrm{f}}}$ (b),
$\mathrm{Net}{\{\mathtt{H},\mathtt{L}, \mathtt{U}\}_{\mathrm{h}1}}$ (c), and
$\mathrm{Net}{\{\mathtt{H},\mathtt{L}, \mathtt{U}\}_{\mathrm{h}2}}$ (d).}
\label{fignetc}
\end{figure}

Furthermore, we propose
$\mathrm{Net}{\{\mathtt{H},\mathtt{L}, \mathtt{U}\}_{\mathrm{d}}}$ and
$\mathrm{Net}{\{\mathtt{H},\mathtt{L}, \mathtt{U}\}_{\mathrm{f}}}$ to fuse all the three modalities at the decision and the feature levels, respectively.  As illustrated in Fig.~\ref{fignetc}~(a) and Fig.~\ref{fignetc}~(b), $\mathrm{Net}{\{\mathtt{H},\mathtt{L}, \mathtt{U}\}_{\mathrm{d}}}$ and
$\mathrm{Net}{\{\mathtt{H},\mathtt{L}, \mathtt{U}\}_{\mathrm{f}}}$ are both composed of
 $\mathrm{Net}{\{\mathtt{H}\}}$,  $\mathrm{Net}{\{\mathtt{L}\}}$, $\mathrm{Net}{\{\mathtt{U}\}}$, and $\mathrm{Net}{\{\mathrm{Fus}1\}}$.
The difference between  $\mathrm{Net}{\{\mathtt{H},\mathtt{L}, \mathtt{U}\}_{\mathrm{d}}}$ and
$\mathrm{Net}{\{\mathtt{H},\mathtt{L}, \mathtt{U}\}_{\mathrm{f}}}$
is that $\mathrm{Net}{\{\mathtt{H},\mathtt{L}, \mathtt{U}\}_{\mathrm{f}}}$ concatenates all the hidden layer outputs rather than the network outputs of $\mathrm{Net}{\{\mathtt{H}\}}$, $\mathrm{Net}{\{\mathtt{L}\}}$, and  $\mathrm{Net}{\{\mathtt{U}\}}$.
Moreover, we  propose
$\mathrm{Net}{\{\mathtt{H},\mathtt{L}, \mathtt{U}\}_{\mathrm{h1}}}$ and
$\mathrm{Net}{\{\mathtt{H},\mathtt{L}, \mathtt{U}\}_{\mathrm{h2}}}$ to fuse all the three modalities at the hybrid level.
As depicted in Fig.~\ref{fignetc}~(c),  $\mathrm{Net}{\{\mathtt{H},\mathtt{L}, \mathtt{U}\}_{\mathrm{h1}}}$ first uses $\mathrm{Net}{\{\mathrm{Fus}1\}}$ to fuse the  hidden layer outputs of
$\mathrm{Net}{\{\mathtt{H}\}}$  and $\mathrm{Net}{\{\mathtt{U}\}}$
and then uses $\mathrm{Net}{\{\mathrm{Fus}2\}}$ to fuse the network output of
$\mathrm{Net}{\{\mathrm{Fus}1\}}$  and the hidden layer output of $\mathrm{Net}{\{\mathtt{L}\}}$. The only difference
between  $\mathrm{Net}{\{\mathtt{H},\mathtt{L}, \mathtt{U}\}_{\mathrm{h1}}}$ and
$\mathrm{Net}{\{\mathtt{H},\mathtt{L}, \mathtt{U}\}_{\mathrm{h2}}}$ is that $\mathrm{Net}{\{\mathtt{H},\mathtt{L}, \mathtt{U}\}_{\mathrm{h1}}}$ fuses the network outputs of both
$\mathrm{Net}{\{\mathrm{Fus}1\}}$  and  $\mathrm{Net}{\{\mathtt{L}\}}$ while $\mathrm{Net}{\{\mathtt{H},\mathtt{L}, \mathtt{U}\}_{\mathrm{h2}}}$ fuses the hidden layer outputs of both
$\mathrm{Net}{\{\mathrm{Fus}1\}}$  and  $\mathrm{Net}{\{\mathtt{L}\}}$.
It should be mentioned that we choose to first fuse
$\mathrm{Net}{\{\mathtt{H}\}}$  and $\mathrm{Net}{\{\mathtt{U}\}}$ at the feature level  because  $\mathrm{Net}{\{\mathtt{H},\mathtt{U}\}_{\textrm{f}}}$ outperforms other proposed two-modality based networks, as will be shown in the later simulation section. This indicates that the fusion of
$\mathord{\buildrel{\lower3pt\hbox{$\scriptscriptstyle\leftarrow$}}
\over {\bm h}} (f_{\textrm{D}})$ and
  $\bm h(f_{\textrm{U}} )$  provides stronger complementarity and therefore would be more suitable to be fused earlier. Note that the design and the testing for DML are not isolated but interoperable, which means that we need the testing results to guide the design of network. In other words, the excellent capability and  flexibility of DML come at the cost of design complexity.

\textbf{\emph{Remark 1:}} The channel prediction based on the three modalities, i.e., $\mathord{\buildrel{\lower3pt\hbox{$\scriptscriptstyle\leftarrow$}}
\over {\bm h}} (f_{\textrm{D}})$,
${\bm D_{{\left(x,y,z\right)} }}$, and  $\bm h(f_{\textrm{U}} )$, can also be referred as the channel extrapolation information across the time, space, and frequency domains. The three-modality based networks in Fig.~\ref{fignetc} jointly exploit the complementarity of  the time-space-frequency information  to improve the performance of  the channel extrapolation.
\subsubsection{Feedback link is available}\label{secbsyes}
In this scenario,
we need to investigate which modality would be  more efficient to
 be fed back to BS under given feedback overhead. When the length of the vector to be fed back, denoted by $T_{\mathrm{fb}}$, is greater than the number of BS antennas $M$, it is obvious that we should directly feed back the downlink channel rather than the received signal $\bm r$.  When  $T_{\mathrm{fb}}$ is smaller than $M$, we respectively try  various fusion networks for
$\{\bm r,\bm s\}$ and $\mathord{\buildrel{\lower3pt\hbox{$\scriptscriptstyle\smile$}}
\over  {\bm h}}(f_{\textrm{D}})$ and then present the networks with the best performance in the following.

\begin{figure*}[!t]
\centering
\includegraphics[width=165 mm]{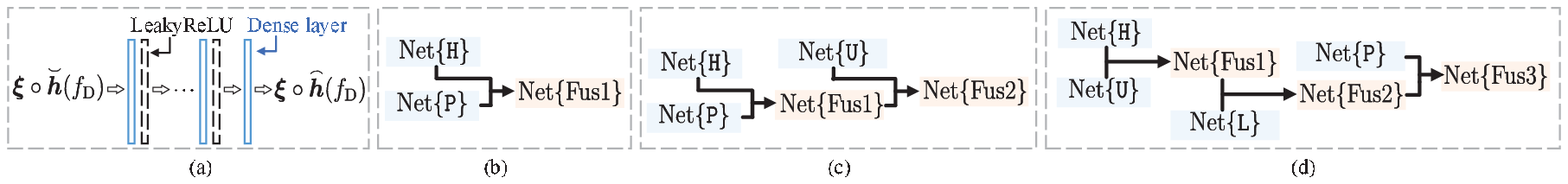}
\caption{ The network structures of $\mathrm{Net}{\{\mathtt{P}\}}$ (a), $\mathrm{Net}{\{\mathtt{P},\mathtt{H}\}}$ (b), $\mathrm{Net}{\{\mathtt{P},\mathtt{H},\mathtt{U}\}}$ (c), and $\mathrm{Net}{\{\mathtt{P},\mathtt{H},\mathtt{L},\mathtt{U}\}}$ (d).}
\label{figbsfeed}
\end{figure*}

\begin{figure*}[!t]
\centering
\includegraphics[width=165 mm]{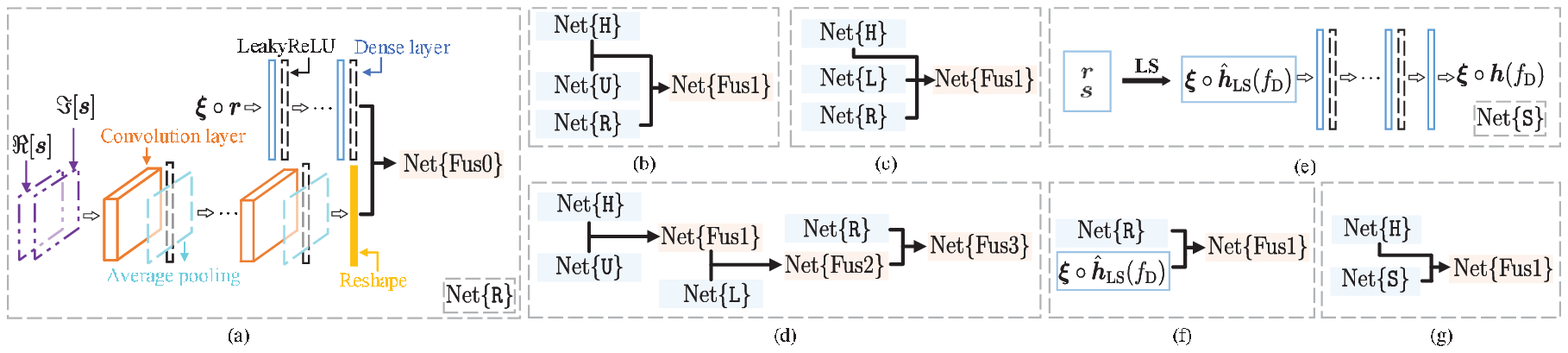}
\caption{The network structures  of  $\mathrm{Net}{\{\mathtt{R}\}}$ (a), $\mathrm{Net}{\{\mathtt{R},\mathtt{H},\mathtt{U}\}}$ (b),
$\mathrm{Net}{\{\mathtt{R},\mathtt{H},\mathtt{L}\}}$ (c), $\mathrm{Net}{\{\mathtt{R},\mathtt{H},\mathtt{L},\mathtt{U}\}}$  (d), $\mathrm{Net}{\{\mathtt{S}\}}$ (e),
$\mathrm{Net}{\{\mathtt{S},\mathtt{R}\}}$ (f), and
$\mathrm{Net}{\{\mathtt{S},\mathtt{H}\}}$ (g). }
\label{figbsp}
\end{figure*}

We first consider the case that $\mathord{\buildrel{\lower3pt\hbox{$\scriptscriptstyle\smile$}}
\over  {\bm h}}(f_{\textrm{D}})$ is fed back to  BS.
Obviously, the length of the vector $\mathord{\buildrel{\lower3pt\hbox{$\scriptscriptstyle\frown$}}
\over {\bm h}} (f_{\textrm{D}})$ is $M\!-\!T_{\mathrm{fb}}$.
As shown in Fig.~\ref{figbsfeed}~(a),
we propose $\mathrm{Net}{\{\mathtt{P}\}}$ to predict unknown $\mathord{\buildrel{\lower3pt\hbox{$\scriptscriptstyle\frown$}}
\over {\bm h}} (f_{\textrm{D}})$ based on the known  $\mathord{\buildrel{\lower3pt\hbox{$\scriptscriptstyle\smile$}}
\over  {\bm h}}(f_{\textrm{D}})$, i.e., to learn the mapping $\bm \Upsilon$.
The network structure of $\mathrm{Net}{\{\mathtt{P}\}}$ is the same with $\mathrm{Net}{\{\mathtt{L}\}}$ except that the
 input and the label of $\mathrm{Net}{\{\mathtt{P}\}}$ are $\bm \xi\circ \mathord{\buildrel{\lower3pt\hbox{$\scriptscriptstyle\smile$}}
\over  {\bm h}}(f_{\textrm{D}})$ and
$\bm \xi\circ \mathord{\buildrel{\lower3pt\hbox{$\scriptscriptstyle\frown$}}
\over {\bm h}} (f_{\textrm{D}})$, respectively.
As shown in Fig.~\ref{figbsfeed}~(b), we propose  $\mathrm{Net}{\{\mathtt{P},\mathtt{H}\}}$ to fuse the network output of $\mathrm{Net}{\{\mathtt{P}\}}$ and the hidden layer output of $\mathrm{Net}{\{\mathtt{H}\}}$.
As presented in Fig.~\ref{figbsfeed}~(c), we propose  $\mathrm{Net}{\{\mathtt{P},\mathtt{H},\mathtt{U}\}}$ to fuse the network output of $\mathrm{Net}{\{\mathtt{P},\mathtt{H}\}}$ and the hidden layer output of $\mathrm{Net}{\{\mathtt{U}\}}$.
As illustrated in Fig.~\ref{figbsfeed}~(d), we propose  $\mathrm{Net}{\{\mathtt{P},\mathtt{H},\mathtt{U},\mathtt{L}\}}$  to fuse the network outputs of both $\mathrm{Net}{\{\mathtt{H},\mathtt{L}, \mathtt{U}\}_{\mathrm{h}2}}$ and $\mathrm{Net}{\{\mathtt{P}\}}$.

Then, we  consider the case that $\bm r$ is fed back to  BS. Since the length of the feedback vector $\bm r$  is smaller than $M$, i.e., $T_p\!=\!T_{\mathrm{fb}}\!<\!M$,
the LS estimator is not applicable due to the rank deficiency. However, it is feasible  for DNNs to learn the
mapping from $\{\bm r,\bm s\}$ to $\bm h(f_{\textrm{D}})$, and thus
we propose $\mathrm{Net}{\{\mathtt{R}\}}$ to predict
$ {\bm h}(f_{\textrm{D}})$ base on $\{\bm r,\bm s\}$.
As shown in Fig.~\ref{figbsp}~(a), $\bm \xi \circ \bm r$ is the input data of the first dense layer while
$\Re[\bm s]$ and $\Im[\bm s]$ are concatenated at a new axis  as the input data of the first convolution layer.
Each convolution layer is followed by the LeakyReLU and
 the average pooling function. The average pooling functions are added to down-sample the data stream and avoid overfitting \cite{SUN201796}.
After reshaping the output of the last convolution layer, we use $\mathrm{Net}{\{\mathrm{Fus}0\}}$ to fuse the two data streams from the modalities $\bm r$ and $\bm s$. The label of $\mathrm{Net}{\{\mathtt{R}\}}$ is  $\bm \xi\circ  {\bm h}(f_{\textrm{D}})$.
Moreover, we propose $\mathrm{Net}{\{\mathtt{R},\mathtt{H},\mathtt{U}\}}$ to fuse the network output of $\mathrm{Net}{\{\mathtt{R}\}}$  with the hidden layer outputs of both $\mathrm{Net}{\{\mathtt{H}\}}$ and $\mathrm{Net}{\{\mathtt{U}\}}$, as depicted in Fig.~\ref{figbsp}~(b).
We also propose $\mathrm{Net}{\{\mathtt{R},\mathtt{H},\mathtt{L}\}}$ to fuse the hidden layer output of $\mathrm{Net}{\{\mathtt{H}\}}$ with
 the network outputs of both $\mathrm{Net}{\{\mathtt{R}\}}$  and $\mathrm{Net}{\{\mathtt{L}\}}$, as shown in Fig.~\ref{figbsp}~(c).
As illustrated in Fig.~\ref{figbsp}~(d), we propose  $\mathrm{Net}{\{\mathtt{R},\mathtt{H},\mathtt{L},\mathtt{U}\}}$  to fuse the network outputs of both $\mathrm{Net}{\{\mathtt{H},\mathtt{L}, \mathtt{U}\}_{\mathrm{h}2}}$ and $\mathrm{Net}{\{\mathtt{R}\}}$.

\textbf{\emph{Remark 2:}} All the networks proposed in Section~\ref{secbs} can be easily extended to other problems like beam prediction and antenna selection. Specifically, by replacing  the labels of all these networks  $\bm \xi\circ  {\bm h}(f_{\textrm{D}})$ with the optimal beam vectors, the proposed architectures can  handle the beam prediction at the BS side. Besides, the proposed architectures can  deal with antenna selection by replacing  the labels of all these networks with the optimal selection vectors.
It is worth mentioning that these variant architectures  do not require  prefect downlink channels, which can significantly reduce the cost resulted from  downlink channel prediction.

\subsection{Architecture Designs at the User Side}\label{secuse}
We consider three different scenarios for  downlink channel prediction at the user side, i.e., pilots being unavailable, insufficient or sufficient.
\subsubsection{Pilots are unavailable}
In this scenario, available modalities are the previous downlink channels $\mathord{\buildrel{\lower3pt\hbox{$\scriptscriptstyle\leftarrow$}}
\over {\bm h}} (f_{\textrm{D}})$ and the user location
${\bm D_{{\left(x,y,z\right)} }}$. As described in Section~\ref{secbsno}, we can use  $\mathrm{Net}{\{\mathtt{H}\}}$, $\mathrm{Net}{\{\mathtt{L}\}}$,   $\mathrm{Net}{\{\mathtt{H},\mathtt{L}\}_{\mathrm{d}}}$
and $\mathrm{Net}{\{\mathtt{H},\mathtt{L}\}_{\mathrm{f}}}$ to predict the downlink channels.

\subsubsection{Pilots are  insufficient}
In this scenario,
 available modalities are $\mathord{\buildrel{\lower3pt\hbox{$\scriptscriptstyle\leftarrow$}}
\over {\bm h}} (f_{\textrm{D}})$,
${\bm D_{{\left(x,y,z\right)} }}$,   $\{\bm r,\bm s\}$, and $\mathord{\buildrel{\lower3pt\hbox{$\scriptscriptstyle\smile$}}
\over  {\bm h}}(f_{\textrm{D}})$. As described in Section~\ref{secbsyes}, we can use
$\mathrm{Net}{\{\mathtt{R}\}}$, $\mathrm{Net}{\{\mathtt{P}, \mathtt{H}\}}$ and
$\mathrm{Net}{\{\mathtt{R},\mathtt{H},\mathtt{L}\}}$ to predict the downlink channels.

\subsubsection{Pilots are sufficient}\label{secusesuf}
When pilots are sufficient,  the LS estimator can be used to estimate the downlink channel.
Inspired by
\cite{8672767} and \cite{8509622}, we propose $\mathrm{Net}{\{\mathtt{S}\}}$, consisting of several dense layers and
LeakyReLU functions, to predict $\bm h(f_{\textrm{D}} )$ based on
the LS estimate of downlink channel  $\hat{\bm h}_{\mathrm{LS}}(f_{\textrm{D}})$, as  illustrated in Fig.~\ref{figbsp}~(e). The input and the label of
$\mathrm{Net}{\{\mathtt{S}\}}$  are $\bm \xi\circ \hat{\bm h}_{\mathrm{LS}}(f_{\textrm{D}})$ and  $\bm \xi\circ\bm h(f_{\textrm{D}} )$, respectively.
It should be emphasized that even when LS estimates have been obtained,  the available modalities, i.e.,  $\mathord{\buildrel{\lower3pt\hbox{$\scriptscriptstyle\leftarrow$}}
\over {\bm h}} (f_{\textrm{D}})$
 and  $\{\bm r,\bm s\}$, could also be utilized to enhance the accuracy.
Moreover, we propose the $\mathrm{Net}{\{\mathtt{S},\mathtt{R}\}}$, as shown in  Fig.~\ref{figbsp}~(f), where the network input of $\mathrm{Net}{\{\mathtt{S}\}}$ and the network output of $\mathrm{Net}{\{\mathtt{R}\}}$ are fused by $\mathrm{Net}{\{\mathrm{Fus}1\}}$. We also propose the $\mathrm{Net}{\{\mathtt{S},\mathtt{H}\}}$ to fuse the network output of $\mathrm{Net}{\{\mathtt{S}\}}$ and the hidden layer output of $\mathrm{Net}{\{\mathtt{H}\}}$, as displayed in Fig.~\ref{figbsp}~(g).

\textbf{\emph{Remark 3:}} The networks $\mathrm{Net}{\{\mathtt{R}\}}$, $\mathrm{Net}{\{\mathtt{R},\mathtt{H},\mathtt{L}\}}$, and $\mathrm{Net}{\{\mathtt{S},\mathtt{R}\}}$  can be easily  extended to data detection.
One simple way inspired by \cite{8052521} is to first divide the transmitted signals into pilots and data signals. Then, the pilots are fed to the network as depicted in Fig.~\ref{figbsp} while the data signals are adopted as the training labels.
In this way, we do not need to collect the downlink channels as training labels, which can significantly reduce the cost for label collection.

\subsection{Training Steps and Complexity Analyses}
The detailed training steps of all the proposed fusion networks are given as follows\footnote{
The number of  training iterations often increases as the network size increases.
In order to  accelerate the off-line training,
we choose to train the elementary networks in advance and then fix these parameters. In fact, all the proposed fusion networks could also be trained end-to-end with its parameters initialized by  the elementary networks.}: 
\begin{enumerate}
  \item  Train the  elementary networks, e.g.,
$\mathrm{Net}{\{\mathtt{H}\}}$, $\mathrm{Net}{\{\mathtt{L}\}}$,  $\mathrm{Net}{\{\mathtt{U}\}}$,
$\mathrm{Net}{\{\mathtt{P}\}}$,
$\mathrm{Net}{\{\mathtt{R}\}}$,
and $\mathrm{Net}{\{\mathtt{S}\}}$,
 independently to minimize the loss between  its output and the label $\bm \xi\circ  {\bm h}(f_{\textrm{D}})$ until their loss functions converge, and then fix these network parameters;
  \item Train $\mathrm{Net}{\{\mathrm{Fus}1\}}$  to minimize the loss between  its output and the label $\bm \xi\circ  {\bm h}(f_{\textrm{D}})$ until its loss function converges, and then fix its network parameters; \label{notenumber}
  \item Following step \ref{notenumber}), train $\mathrm{Net}{\{\mathrm{Fus}2\}}$  and $\mathrm{Net}{\{\mathrm{Fus}3\}}$ successively until their loss function converges and then fix their network parameters successively.
\end{enumerate}
The required number of floating point operations  is used as the metric of complexity. Then,
the  complexity of the dense layer is $O(n_{\textrm{in}}n_{\textrm{out}})$, where $n_{\textrm{in}}$ and $n_{\textrm{out}}$ are respectively the input and the output data sizes. The  complexity of the convolution layer is $O(n_{\textrm{in}}n_{\textrm{Flit}}n_{\textrm{Flit,in}}n_{\textrm{Flit,out}})$, where $n_{\textrm{Flit}}$ is the filter size while $n_{\textrm{Flit,in}}$ and $n_{\textrm{Flit,out}}$ are respectively the input and the output numbers of  filters.
The  complexity of the LSTM layer is $O(4T_{\mathrm{unit}}n_{\textrm{in}}n_{\textrm{out}})$.
Obviously, the computation cost increases as the number of modalities  increases. The balance between the  cost and  the performance should be considered in practical applications.

\section{Simulation Results}\label{secsimu}
In this section, we will first present the simulation scenario and default network parameters.
Then, the  performance of the proposed networks
will be evaluated and analyzed.

\begin{figure}[!t]
\centering
\includegraphics[width=70mm]{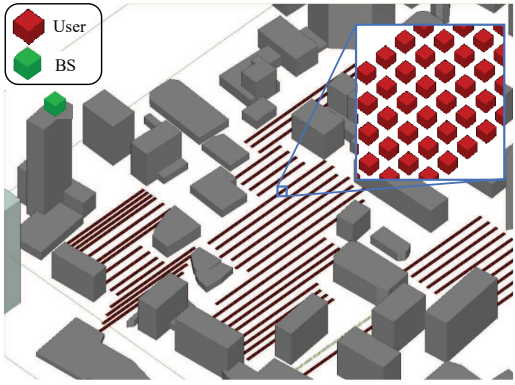}
\caption{A partial view of the ray-tracing scenario. The green little box represents the BS antennas.  The red little box represents the possible location of the user antenna and the distance between adjacent red little boxes is 1 m. The red line is consistent with the y-axis.
This ray-tracing simulator shoots thousands of rays in all directions from the transmitter and records the strongest 25 paths that reach the receiver \cite{alkhateeb2019deepmimo}.}
\label{figcase}
\end{figure}

\begin{table}[!t]\footnotesize
\centering
\caption{Default Parameters for the proposed networks}
\label{tabpara}
\begin{tabular}{|c|c|c|c|}
\hline
Modality & Network  & Structure parameter  & Learning rate \\
\hline
$\mathord{\buildrel{\lower3pt\hbox{$\scriptscriptstyle\leftarrow$}}
\over {\bm h}} (f_{\textrm{D}})$ & $\mathrm{Net}{\{\mathtt{H}\}}$ &   $T_u=3$, LSTM: 256,256 & 5e-4 \\
\hline
${\bm D_{{\left(x,y,z\right)} }}$ & $\mathrm{Net}{\{\mathtt{L}\}}$ &  Dense: 256,256,256,256 & 1e-4 \\
\hline
 $\bm h(f_{\textrm{U}} )$& $\mathrm{Net}{\{\mathtt{U}\}}$ &  Dense: 256,256,256 & 1e-3 \\
\hline
$\mathord{\buildrel{\lower3pt\hbox{$\scriptscriptstyle\smile$}}
\over  {\bm h}}(f_{\textrm{D}}) $ & $\mathrm{Net}{\{\mathtt{P}\}}$ &  Dense: 256,256,256 & 5e-4 \\
\hline
$\{\bm r,\bm s\}$ & $\mathrm{Net}{\{\mathtt{R}\}}$ &
\tabincell{c}{Dense: 256,128,128\\   Filter number: 16, 32, 8\\ Filter: (5, 5)}& 5e-4 \\
\hline
$ \hat{\bm h}_{\mathrm{LS}}(f_{\textrm{D}})$ & $\mathrm{Net}{\{\mathtt{S}\}}$ &  Dense: 256 & 5e-4 \\
\hline
 -& \tabincell{c}{$\mathrm{Net}{\{\mathrm{Fus}i\}}$\\
{i=0,1,2,3}\\}
 &  Dense: 512,512,256 &  5e-4  \\
\hline
\end{tabular}
\end{table}

\begin{figure}[!t]
\centering
\includegraphics[width=80 mm]{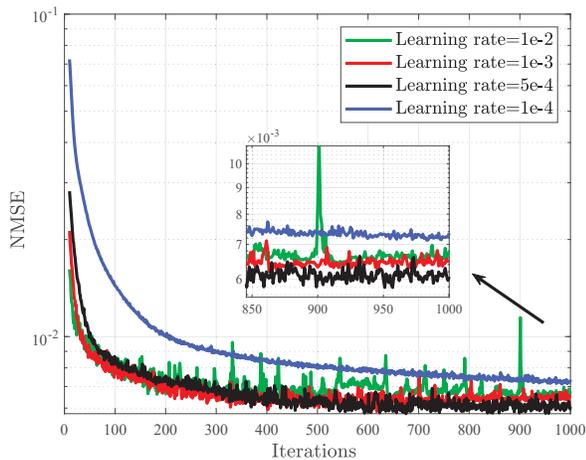}
\caption{The NMSE performance of $\mathrm{Net}{\{\mathtt{H}\}}$ versus different learning rates.}
\label{figlr}
\end{figure}

\subsection{Simulation Setup}
\textbf{Dataset Generation:} In the simulations, we consider the outdoor massive MIMO scenario that is constructed based on   the accurate 3D ray-tracing simulator  Wireless InSite \cite{alkhateeb2019deepmimo}.
Unlike conventional statistical channel generation methods,
the 3D ray-tracing simulator can capture the dependence of channels on the environment geometry/materials and transmitter/receiver locations,  and therefore can provide more reliable datasets for training and testing.
The scenario comprises  one BS  and massive randomly distributed  user antennas and each BS is  equipped with 64 antennas. The scenario covers an area of $500\times500$ square metres.
A partial view of the ray-tracing scenario is illustrated in  Fig.~\ref{figcase}.
The uplink and downlink frequencies are set to be 2.50 GHz and 2.62 GHz, respectively.
Based on the environment setup, the 3D ray-tracing simulator first outputs the uplink channel parameters, the downlink channel parameters,  and the location ${{\bm D_{{\left(x,y,z\right)} }}}$ for each user.
Then,
we can obtain $\bm h(f_{\textrm{U}} )$ and $\bm h(f_{\textrm{D}} )$ for each user by using
Eq.~\eqref{equchannel} and the channel parameters from the 3D ray-tracing simulator.
With Eq.~\eqref{equreceive}, we can generate the pilots and the received signals
$\{\bm r,\bm s\}$ based on $\bm h(f_{\textrm{D}})$.
Assuming that the previous downlink channels  are the channels of the user at adjacent positions, and the users move along the y-axis\footnote{One similar example of using ray-tracing  to model
the time correlations can be found in  \cite{8580703}. For more details about how to generate channels using Wireless InSite, please refer to the paper  \cite{alkhateeb2019deepmimo} and the codes \cite{code2}.}. Then,  $\mathord{\buildrel{\lower3pt\hbox{$\scriptscriptstyle\leftarrow$}}
\over {\bm h}} (f_{\textrm{D}})$  can be obtained by collecting channels at adjacent $T_{\textrm{unit}}$ positions.
Partial downlink channel $\mathord{\buildrel{\lower3pt\hbox{$\scriptscriptstyle\smile$}}
\over  {\bm h}}(f_{\textrm{D}})$  can be obtained
by selecting $M_{\mathrm{fb}}$ elements out of  $\bm h(f_{\textrm{D}})$, and then the rest $M-M_{\mathrm{fb}}$ elements constitute the vector $\mathord{\buildrel{\lower3pt\hbox{$\scriptscriptstyle\frown$}}
\over  {\bm h}}(f_{\textrm{D}})$.
After obtaining all sample pairs, we randomly select 9000 samples from the sample pairs as the training dataset, and select 1000 samples from the rest of sample pairs as the testing dataset.
Since the perfect channels are not available in practical situation, unless otherwise specified, all the sample pairs in the datasets are estimated by the LMMSE algorithm  \cite{1597555} when the signal-to-noise ratio (SNR) is 25 dB. 

\textbf{Adopted Neural Networks:} Unless otherwise specified, the  parameters of the proposed networks are given  in Tab.~\ref{tabpara}, where
``LSTM: 256, 256'' means that the hidden layers in $\mathrm{Net}{\{\mathtt{H}\}}$ consist of two LSTM layers, and each hidden layer  has 256  units.
The numbers of  units in the input and the output layers for all the proposed networks are  consistent with the lengths of the input and the output data vectors, and thus are omitted in Tab.~\ref{tabpara}. We choose the output of the middle hidden layer as the hidden layer output of the networks. The batch size $V$ of all proposed networks is 128.
To facilitate the understanding, we also list the involved modalities for the elementary networks in Tab.~\ref{tabpara}. For the proposed fusion networks,  ${\{\cdot\}_{\mathrm{f}}}$, ${\{\cdot\}_{\mathrm{d}}}$, ${\{\cdot\}_{\mathrm{h}}}$ respectively represent the feature, the decision, and the hybrid fusions. For natation simplicity, some hybrid fusion networks have omitted the notation $\mathrm{h}$, i.e., $\mathrm{Net}{\{\mathtt{P}, \cdot\}}$, $\mathrm{Net}{\{\mathtt{R}, \cdot\}}$, and $\mathrm{Net}{\{\mathtt{S}, \cdot\}}$.

 Normalized mean-squared-error (NMSE) is used to measure the prediction accuracy, and is defined as
\[{\rm{NMSE}} = E \left[\left\|{\bm h}_{\textrm{D}}-\hat{\bm h}_{\textrm{D}}\right\|_{2}^{2}/\left\|{\bm h}_{\textrm{D}}\right\|_{2}^{2}\right],\] where $\hat{\bm h}_{D}$ and  ${\bm h}_{D} $  represent the estimated and the true downlink channels, respectively.
 Note that all the parameters in Tab.~\ref{tabpara} are basically selected by trails and errors such that these algorithms perform well. Take the learning rate selection of $\mathrm{Net}{\{\mathtt{H}\}}$ as an example. As illustrated in  Fig.~\ref{figlr}, the accuracy stability improves as the learning rate decreases. In particular,  $\mathrm{Net}{\{\mathtt{H}\}}$  can achieve the highest accuracy when the learning rate is 5e-4.

\begin{figure}[!t]
\centering
\includegraphics[width=80 mm]{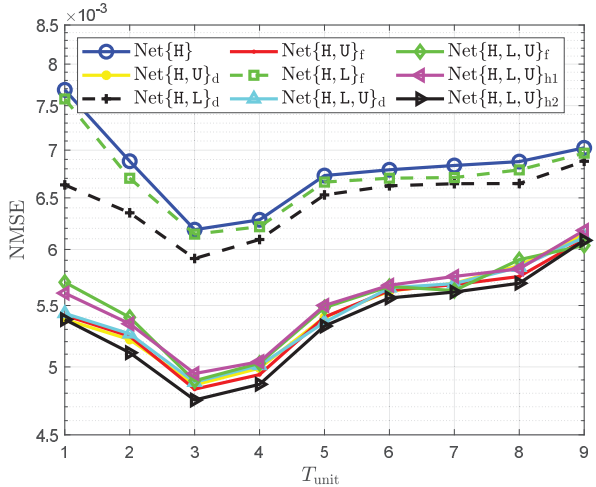}
\caption{The NMSE performance of previous downlink channel related networks versus  $T_{\mathrm{unit}}$. }
\label{figsimulstm}
\end{figure}

\begin{figure*}[!t]
\centering
\includegraphics[width=135 mm]{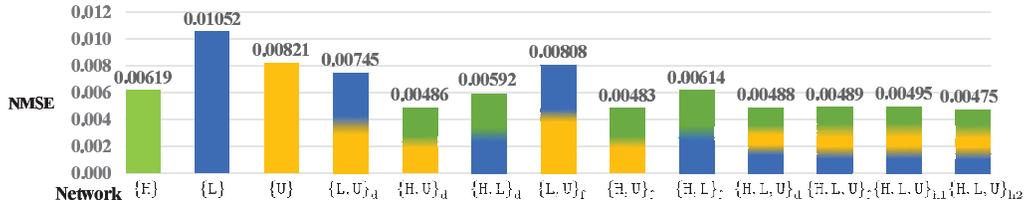}
\caption{The NMSE performance of the networks that are applicable to BS without feedback link. }
\label{figbsno}
\end{figure*}

\subsection{BS Side}
Fig.~\ref{figsimulstm} displays the  NMSE performance of the previous downlink channel related networks
 versus  $T_{\mathrm{unit}}$.
A  larger  $T_{\mathrm{unit}}$ means that $\mathrm{Net}{\{\mathtt{H}\}}$ would learn time correlation of downlink channels from
previous downlink channels in longer time periods.
It can be observed that the performance of  all these networks first improves and then degrades  as $T_{\mathrm{unit}}$ increases. This is because that the time correlations among the current channel and  previous downlink channels can help to improve the performance of the proposed networks while  the time correlations tend to vanish as $T_{\mathrm{unit}}$  increases.
Furthermore, $\mathrm{Net}{\{\mathtt{H}\}}$ consistently achieves the worst performance, and $\mathrm{Net}{\{\mathtt{H},\mathtt{L}, \mathtt{U}\}_{\mathrm{h}2}}$ consistently outperforms  other networks. This implies that the modalities
${\bm D_{{\left(x,y,z\right)} }}$ and $\bm h(f_{\textrm{U}} )$ do provide complementary information for $\mathrm{Net}{\{\mathtt{H}\}}$.
We will set $T_{\mathrm{unit}}$ to be 3 for better performance in the following simulations.

Fig.~\ref{figbsno} shows the NMSE performance of all  networks that are applicable to BS without  feedback link, as discussed in Section~\ref{secbsno}. As shown in Fig.~\ref{figbsno}, all the two-modalities fused networks outperform the corresponding two single-modality networks, which implies that any two of the three modalities, i.e., $\mathord{\buildrel{\lower3pt\hbox{$\scriptscriptstyle\leftarrow$}}
\over {\bm h}} (f_{\textrm{D}})$,
${\bm D_{{\left(x,y,z\right)} }}$ and $\bm h(f_{\textrm{U}} )$, can provide complementary information, thus enhancing the prediction accuracy.
In particular, although  $\mathrm{Net}{\{\mathtt{L}\}}$ has worse performance than both $\mathrm{Net}{\{\mathtt{H}\}}$ and $\mathrm{Net}{\{\mathtt{U}\}}$, the four two-modalities fused networks, i.e., $\mathrm{Net}{\{\mathtt{H},\mathtt{L}\}_{\mathrm{d}}}$, $\mathrm{Net}{\{\mathtt{H},\mathtt{L}\}_{\mathrm{f}}}$, $\mathrm{Net}{\{\mathtt{L},\mathtt{U}\}_{\mathrm{d}}}$, and $\mathrm{Net}{\{\mathtt{L},\mathtt{U}\}_{\mathrm{f}}}$, all have better performance than  both $\mathrm{Net}{\{\mathtt{H}\}}$ and $\mathrm{Net}{\{\mathtt{U}\}}$.
Besides, we notice that $\mathrm{Net}{\{\mathtt{H},\mathtt{U}\}_{\mathrm{f}}}$ has the best performance among two-modalities fused networks, and the hybrid fusion network $\mathrm{Net}{\{\mathtt{H},\mathtt{L}, \mathtt{U}\}_{\mathrm{h}2}}$ outperforms  other than   feature or decision  fusion networks. In fact, the structure of
$\mathrm{Net}{\{\mathtt{H},\mathtt{L}, \mathtt{U}\}_{\mathrm{h}2}}$ is inspired by that of $\mathrm{Net}{\{\mathtt{H},\mathtt{U}\}_{\mathrm{f}}}$. More specifically, since $\mathrm{Net}{\{\mathtt{H},\mathtt{U}\}_{\mathrm{f}}}$  outperforms other two-modalities fused networks, we choose to preferentially fuse $\mathord{\buildrel{\lower3pt\hbox{$\scriptscriptstyle\leftarrow$}}
\over {\bm h}} (f_{\textrm{D}})$ and $\bm h(f_{\textrm{U}} )$ at the feature level.
Following the design of $\mathrm{Net}{\{\mathtt{H},\mathtt{L}, \mathtt{U}\}_{\mathrm{h}2}}$, we develop the network structures of other fusion networks and only present these with best performance, i.e.,  $\mathrm{Net}{\{\mathtt{P}, \cdot\}}$, $\mathrm{Net}{\{\mathtt{R}, \cdot\}}$, and $\mathrm{Net}{\{\mathtt{S}, \cdot\}}$, which are exactly all
 hybrid fusion network. In summary, compared with the feature and the decision fusion networks, hybrid fusion networks  can often achieve better performance but require higher design complexity. Moreover, the superiority of the feature and the decision fusion depends on the specific task.

\begin{figure}[!t]
\centering
\includegraphics[width=80 mm]{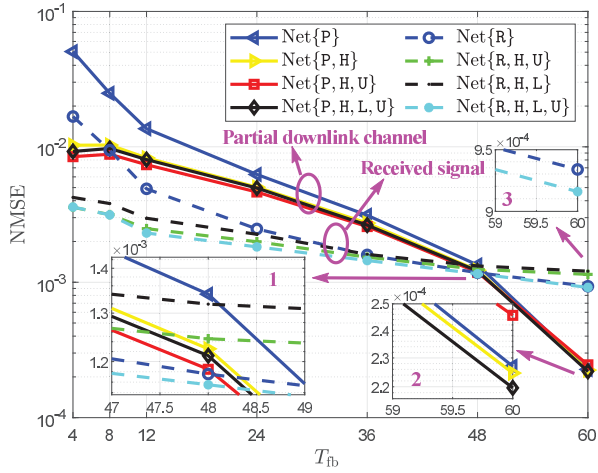}
\caption{The NMSE performance of all the networks that are applicable to BS with feedback link. The networks are trained separately for each value of $T_{\mathrm{bf}}$.}
\label{figfeedback}
\end{figure}

Fig.~\ref{figfeedback} compares the NMSE performance of all the networks that are applicable to BS with  feedback link, as discussed in Section~\ref{secbsyes}.
 As shown in Fig.~\ref{figfeedback}, the performance of all the proposed networks improves when the feedback length $T_{\mathrm{fb}}$ increases.
  When $T_{\mathrm{fb}}$ is small, the received signal based networks (solid lines) outperform  partial downlink channel based networks (dotted  lines). This is because the correlations between the  received signals and downlink channels are stronger than  the correlations between the channels of different antennas. Since the estimate error of the unknown channels becomes insignificant as  $T_{\mathrm{fb}}$ increases, the prediction accuracy of partial downlink channel based networks would outperform the received signal based networks. Consider a extreme case, where $T_{\mathrm{fb}}$ equals 64. Then, the NMSE  of partial downlink channel based networks would be 0.
Furthermore, it can be observed from the second enlarge picture that
$\mathrm{Net}{\{\mathtt{P},\mathtt{H}\}}$ and $\mathrm{Net}{\{\mathtt{P},\mathtt{H},\mathtt{L},\mathtt{U}\}}$
consistently outperform  $\mathrm{Net}{\{\mathtt{P}\}}$ while the gaps between the three networks all degrade as $T_{\mathrm{fb}}$ increases.
This indicates that when we choose to feed partial downlink channel back, i.e., $T_{\mathrm{fb}}\!\ge\!48$, we can adopt $\mathrm{Net}{\{\mathtt{P}\}}$ instead of other $\mathord{\buildrel{\lower3pt\hbox{$\scriptscriptstyle\smile$}}
\over  {\bm h}}(f_{\textrm{D}})$ related fusion networks to reduce the training cost, since the gaps between them are negligible.
Moreover, as shown in the third enlarge picture,
$\mathrm{Net}{\{\mathtt{R},\mathtt{H},\mathtt{L},\mathtt{U}\}}$
consistently outperforms  $\mathrm{Net}{\{\mathtt{R}\}}$, $\mathrm{Net}{\{\mathtt{R},\mathtt{H},\mathtt{U}\}}$, and
$\mathrm{Net}{\{\mathtt{R},\mathtt{H},\mathtt{L}\}}$  while the gap between $\mathrm{Net}{\{\mathtt{R},\mathtt{H},\mathtt{L},\mathtt{U}\}}$ and $\mathrm{Net}{\{\mathtt{R}\}}$ becomes negligible when
 $T_{\mathrm{fb}}$  is larger than 36.  This indicates that we can adopt $\mathrm{Net}{\{\mathtt{R},\mathtt{H},\mathtt{L},\mathtt{U}\}}$  for better prediction accuracy when $T_{\mathrm{fb}}$  is smaller than 36 and adopt $\mathrm{Net}{\{\mathtt{R}\}}$  for lower training cost when  $T_{\mathrm{fb}}$ is greater than 36 and less than 48.
Besides, when the feedback length is low, the extra accuracy gain obtained by using multi-modal data is remarkable, which also demonstrates the significant superiority of DML in FDD massive MIMO systems.

\subsection{User Side}
\begin{figure}[!t]
\centering 
\begin{minipage}[b]{0.48\textwidth} 
\centering 
\includegraphics[width=1\textwidth]{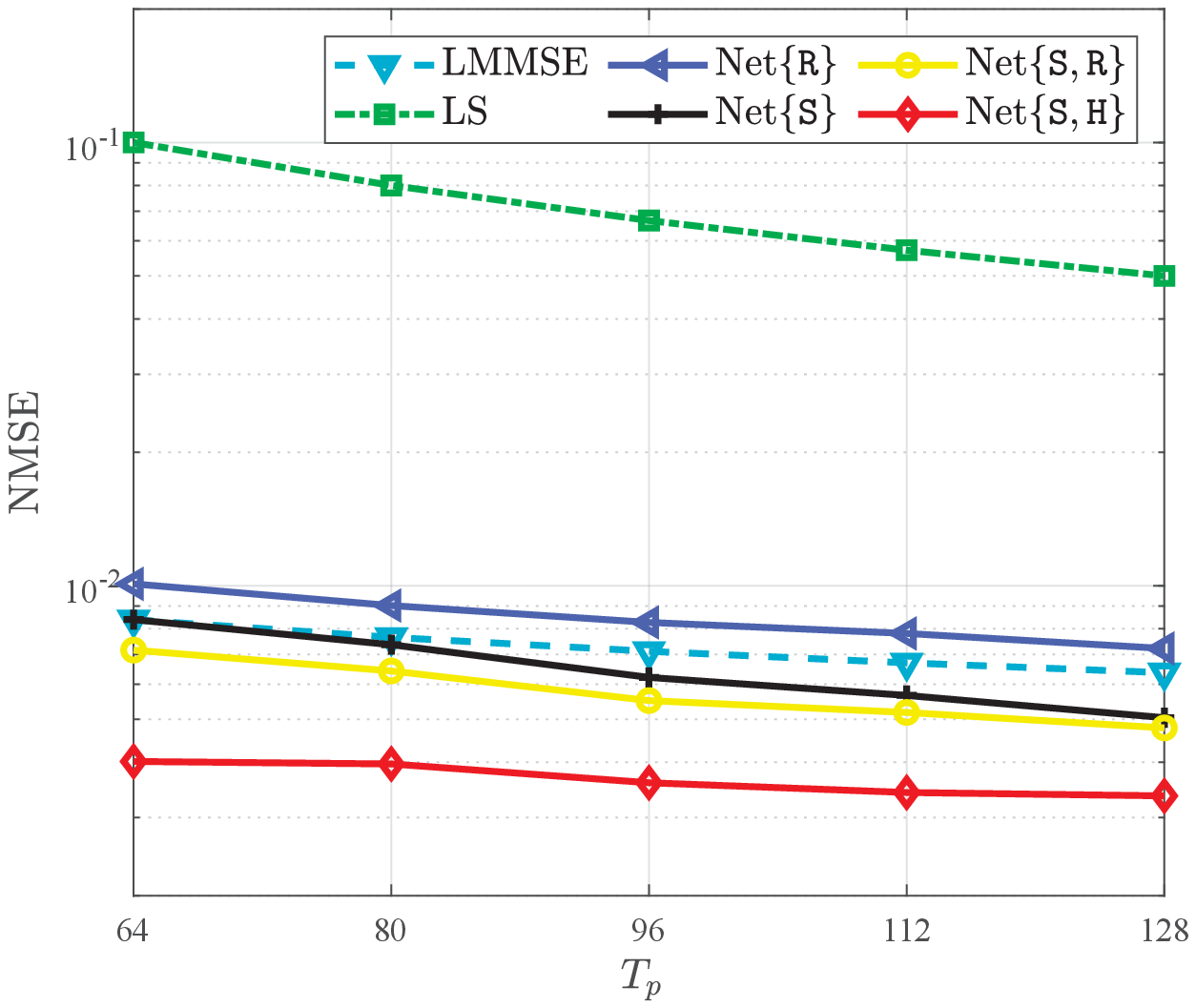} 
\caption{The NMSE performance of the proposed networks   versus the pilot length $T_{p}$. The networks are trained separately for each value of $T_{p}$, and SNR is 10 dB in both the training and the testing stages. }
\label{figls}
\end{minipage}
\begin{minipage}[b]{0.48\textwidth} 
\centering 
\includegraphics[width=1\textwidth]{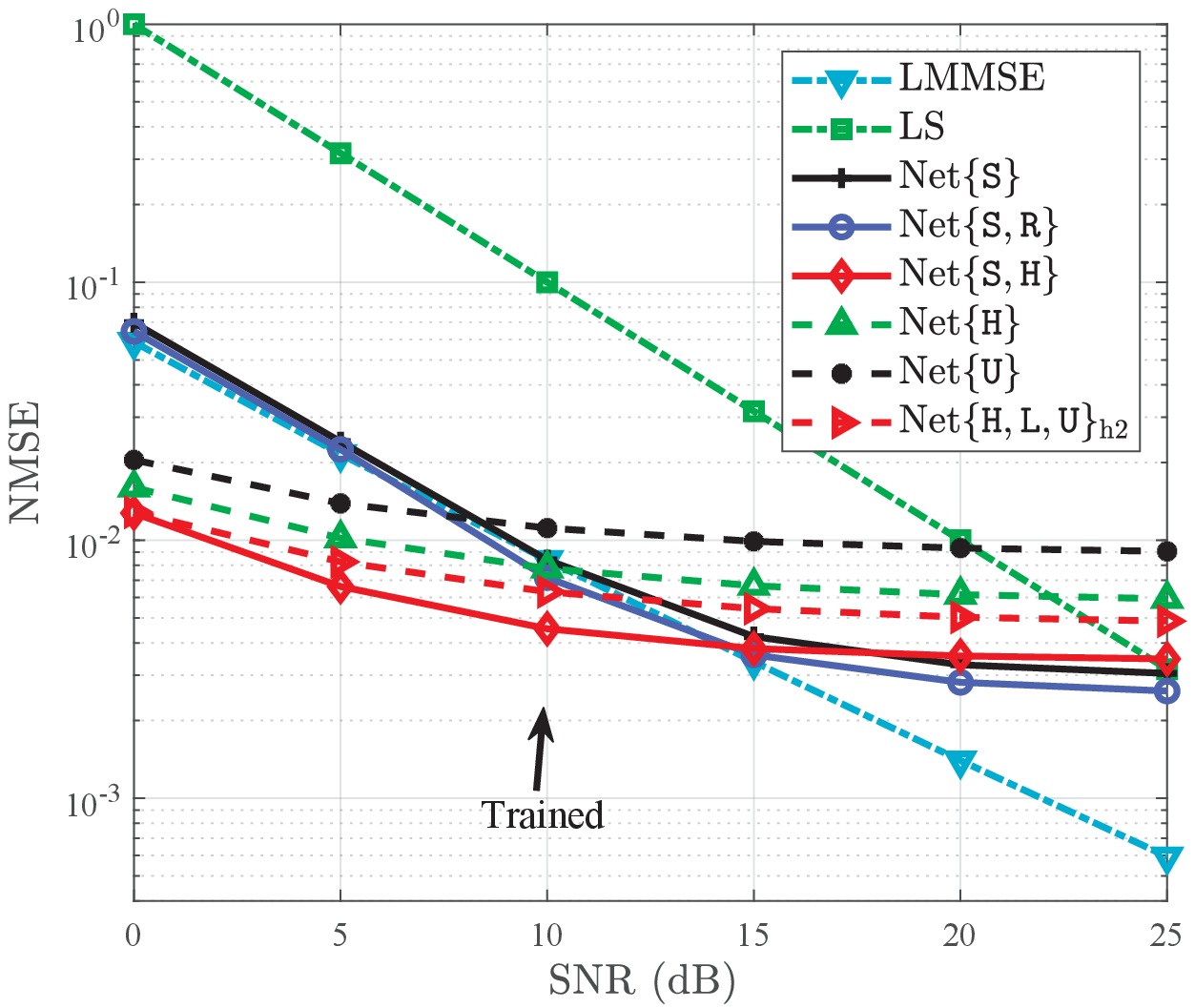}
\caption{The NMSE performance of the proposed networks  versus  SNR, where $T_{p}$ is 64. The networks are trained at SNR=10 dB and tested separately for each value of SNR.}
\label{figsnr}
\label{figsnr}
\end{minipage}
\end{figure}

Fig.~\ref{figls} displays
the NMSE performance of LS, LMMSE, $\mathrm{Net}{\{\mathtt{R}\}}$, $\mathrm{Net}{\{\mathtt{S}\}}$, $\mathrm{Net}{\{\mathtt{S},\mathtt{R}\}}$, and
$\mathrm{Net}{\{\mathtt{S},\mathtt{H}\}}$
versus the pilot length $T_{p}$, where SNR is 10 dB in both the training and the testing stages. The pervious channels involved in $\mathrm{Net}{\{\mathtt{S},\mathtt{H}\}}$  are  estimated by the LMMSE algorithm  when SNR is 10 dB.
As shown in Fig.~\ref{figls}, both $\mathrm{Net}{\{\mathtt{S},\mathtt{H}\}}$ and  $\mathrm{Net}{\{\mathtt{S},\mathtt{R}\}}$   outperform $\mathrm{Net}{\{\mathtt{S}\}}$ and $\mathrm{Net}{\{\mathtt{R}\}}$,  which means that extra modalities can    provide complementary information to improve the prediction accuracy.
Furthermore, $\mathrm{Net}{\{\mathtt{S},\mathtt{H}\}}$ outperforms all other algorithms, including the  LMMSE algorithm, which demonstrates its remarkable effectiveness.

%

\subsection{Impairments in Practical Applications}
To collect off-line training samples,  we can obtain extremely accurate channels by increasing SNR and the  pilot length. However, in the on-line  testing stage, varying SNRs would impair the  prediction accuracy of the proposed networks. Therefore, we investigate the impact of various SNRs on the performance of LS, LMMSE,
$\mathrm{Net}{\{\mathtt{R}\}}$, $\mathrm{Net}{\{\mathtt{S}\}}$,
$\mathrm{Net}{\{\mathtt{S},\mathtt{R}\}}$,
$\mathrm{Net}{\{\mathtt{S},\mathtt{H}\}}$,
$\mathrm{Net}{\{\mathtt{H}\}}$,
$\mathrm{Net}{\{\mathtt{U}\}}$, and
$\mathrm{Net}{\{\mathtt{H},\mathtt{L}, \mathtt{U}\}_{\mathrm{h}2}}$,
where  $T_{p}$ is 64.
Fig.~\ref{figsnr} shows the performance of these networks
versus the SNR in the on-line testing stage, where the networks are trained at SNR=10 dB and tested separately for each value of SNR.
As indicated in Fig.~\ref{figsnr}, $\mathrm{Net}{\{\mathtt{S},\mathtt{H}\}}$  outperforms all other algorithms when SNR is lower than 15 dB. This is because that the estimation based on the pilots and the  received signals   relies on high SNRs while the
prediction based on  pervious downlink channels is more robust over low SNRs. Since the pilots and the  received signals would be more correlated  with the downlink channel as the SNR increases, pilot related networks would outperform  $\mathrm{Net}{\{\mathtt{H}\}}$,
$\mathrm{Net}{\{\mathtt{U}\}}$, and
$\mathrm{Net}{\{\mathtt{H},\mathtt{L}, \mathtt{U}\}_{\mathrm{h}2}}$  in high SNR region.
 Moreover, the remarkable  robustness of $\mathrm{Net}{\{\mathtt{S},\mathtt{H}\}}$ over low SNRs also demonstrates its potential in real applications.

\textbf{\emph{Discussion: }} The design and the testing for DML are interoperable, which is a typical characteristic of data-driven techniques. On the  contrary, conventional communications are model-driven, where the algorithm designs rely on  prerequisite models and their performance is completely predictable and explicable. In this paper, we have provided a heuristic framework for DML based communications, and the theoretical guidance on the framework will be left for future work.
It should be mentioned that the proposed DML based framework can also be used in cell-free or multi-cell massive MIMO systems \cite{8815888,9113273}.

\section{Conclusion}\label{secconcul}
In this paper, we introduced DML into wireless communications to fully exploit the MSI in communication systems such that the system performance  could be improved. We provided  complete descriptions and heuristic analyses of the design choices in DML based wireless communications.
We also proposed several efficient DML based architectures for channel prediction as a case study.
 Simulation results have shown that DML based architectures exhibit significant advantages over one modality based architectures under most cases, which also  demonstrate that
  the proposed framework can effectively exploit the constructive and complementary information of multimodal sensory data to assist the current wireless communications.

\bibliographystyle{IEEEbib}
\bibliography{References}


\end{document}